\documentclass[onecolumn,amsmath,amssymb,12pt,superscriptaddress,nofootinbib,floatfix]{revtex4}
\pdfoutput=1

\usepackage[latin1]{inputenc}
\usepackage[english]{babel}
\usepackage{amssymb}
\usepackage{amsmath}
\usepackage{amsthm}
\usepackage[]{graphicx}
\usepackage[]{subfigure}
\usepackage{tensor}
\usepackage{color}
\usepackage{cancel}
\usepackage{setspace}
\usepackage{fancyhdr}
\usepackage[bookmarks,linktocpage, colorlinks=true, plainpages = false, citecolor = blue,  linkcolor=blue, urlcolor = blue, filecolor = blue]{hyperref}

\def\ah{\hat a}
\def\Nh{\hat N}
\def\phih{\hat\phi}

\def\1p{{(1p)}}

\def\p0{\phi_0}
\def\pnb{p_{NB}}
\def\ptra{p_{\rm trans}(\p0'',\p0')}
\def\ptr{p_{\rm trans}}

\def\be{\begin{equation}}
\def\ee{\end{equation}}
\def\beq{\begin{eqnarray}}
\def\eeq{\end{eqnarray}}

\begin{document}

\title{Quantum Transitions Through Cosmological Singularities}
\author{Sebastian F. Bramberger}
\email{sebastian.bramberger@aei.mpg.de}
\affiliation{Max Planck Institute for Gravitational Physics (Albert Einstein Institute), 14476 Potsdam-Golm, Germany}
\author{Thomas Hertog}
\email{thomas.hertog@kuleuven.be}
\affiliation{Institute for Theoretical Physics, KU Leuven, 3001 Leuven, Belgium}
\author{Jean-Luc Lehners}
\email{jlehners@aei.mpg.de}
\affiliation{Max Planck Institute for Gravitational Physics (Albert Einstein Institute), 14476 Potsdam-Golm, Germany}
\author{Yannick Vreys}
\email{yannick.vreys@kuleuven.be}
\affiliation{Institute for Theoretical Physics, KU Leuven, 3001 Leuven, Belgium}

\begin{abstract}
\vspace{1cm}
\noindent In a quantum theory of cosmology spacetime behaves classically only in limited patches of the configuration space on which the wave function of the universe is defined. Quantum transitions can connect classical evolution in different patches. Working in the saddle point approximation and in minisuperspace we compute quantum transitions connecting inflationary histories across a de Sitter like throat or a singularity. This supplies probabilities for how an inflating universe, when evolved backwards, transitions and branches into an ensemble of histories on the opposite side of a quantum bounce. Generalising our analysis to scalar potentials with negative regions we identify saddle points describing a quantum transition between a classically contracting, crunching ekpyrotic phase and an inflationary universe. 
\end{abstract}

\maketitle
\newpage
	\tableofcontents
	\newpage
%%%%%%%%%%%%%%%%%%%
\section{Introduction}
%%%%%%%%%%%%%%%%%%%

In a universe that is fundamentally quantum mechanical, classical cosmological evolution may not be predicted by the universe's quantum state for all times and in all regions of the configuration space on which the wave function $\Psi$ is defined. Instead one expects histories of the universe to behave classically in limited patches only. Regions of our universe where classical evolution likely breaks down include the high curvature realm of the early universe and the interior of black holes. 

Classical evolution emerges when the quantum probabilities are high for histories with deterministic correlations in time. These quantum probabilities are given by the quantum state and cannot be reliably diagnosed from the classical equations of motion for cosmology any more than this can be done for any other quantum system. This means in particular that classical evolution can break down without the breakdown of the classical equations of motion. An example of this is the universe's evolution near the de Sitter like throat in the inflationary histories in the no-boundary wave function \cite{Hartle:2015bna}. 

Histories of the universe do not simply end when they cease to behave classically. Rather classical evolution is replaced by quantum evolution. In this paper we apply the framework of minisuperspace quantum cosmology to study what happens when classical cosmological evolution breaks down in the early universe, in the context of four dimensional Einstein gravity coupled to a scalar field. In particular, we find saddle point solutions of the Lorentzian path integral for this model that describe transitions connecting two patches where the universe behaves classically, according to Einsteins equations. 

\begin{figure}[ht!]
	\centering
	\includegraphics[width=0.3\textwidth]{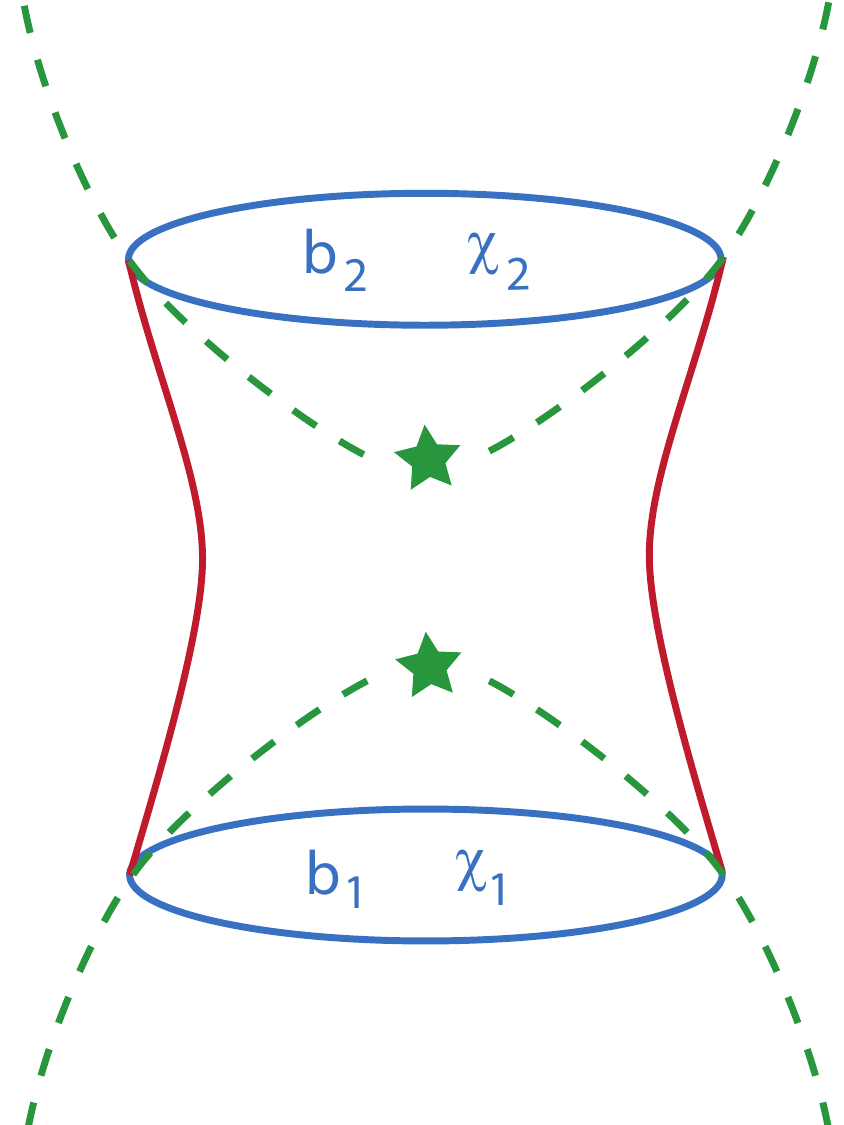}
	\caption{We will study quantum transitions (in red) between real, classical boundaries (in blue), here shown for the situation in which the classical evolution (in green) would lead to singularities.}
	\label{fig:qt}
\end{figure}

The classical behaviour of the boundary configurations means that the saddle points obey boundary conditions corresponding to real values of the scalar field and scale factor on both ends of the transition (cf. Fig. \ref{fig:qt}). The solutions are complex in the interior, however, as is common for gravitational instantons. The action of the saddle points specifies probabilities for quantum transitions through the region of breakdown of classical evolution, thereby connecting a classical history in a given patch with a classical history in another patch. One history typically branches into many histories \cite{Hartle:2015bna}. 

We will apply our method to two qualitatively different cosmological scenarios that are of special interest: quantum transitions between inflationary histories on both ends, and transitions from ekpyrosis to inflation. The classical extrapolation of the histories beyond its domain of validity can produce a curvature singularity in both cases. In inflation the singularity lies in the past, whereas ekpyrotic histories contain a singularity in their future. The quantum transitions we find can thus be viewed as a resolution -- albeit in minisuperspace --  of the cosmological singularity in these models.

Transitions between inflationary histories may be argued to be somewhat academic in that the opposite side of the bounce is unlikely to lead to testable predictions on our side of the bounce. This is because the physical arrows of time point away from the bounce on both sides in all quantum states, such as the no-boundary state \cite{Hartle1983}, that implies perturbations are in their vacuum state near the bounce \cite{Hartle:2011rb}. By contrast, transitions from ekpyrosis to inflation are central to the theory because the arrow of time does not reverse in ekpyrotic cosmology, where the detailed spectral properties of the perturbations on our side generally depend on the conditions before and at the bounce.

%%%%%%%%%%%%%%%%%
%%%%%%%%%%%%%%%%%

The plan of this paper is as follows: we start by describing the general framework that we will work with, namely the semiclassical path integral for quantum gravity (\ref{histories}) in the minisuperspace approximation. We find saddle points of the path integral describing transitions between inflationary histories in Section \ref{inftoinf}. In Section \ref{nbwf-classens}
we use these to connect inflationary histories in the no-boundary state, enabling us to construct probabilities for complete inflationary histories exhibiting a quantum bounce in Section \ref{sec:prob}. In section \ref{section:ekinf} we find transitions between an ekpyrotic contracting phase and an inflationary expanding phase. We conclude in Section \ref{discussion} and provide further technical details in the appendix.

%%%%%%%%%%%%%%%%%%%%%%%%%
\section{Quantum Transitions of the Universe }
\label{histories}

We work in a minisuperspace model in which the Lorentzian four-geometries are homogeneous, isotropic, and spatially closed on the manifold $M={\bf R}\times S^3$. For the matter content we take a single homogeneous scalar field $\phi$ moving in a potential $V(\phi)$. The metrics of homogeneous, isotropic, Lorentzian (quantum mechanical) histories in standard coordinates are  
\be
ds^2 = -\Nh^2(\lambda) d\lambda^2 + \ah^2(\lambda) d\Omega^2_3 \ ,
\label{homoiso}
\ee
where $d\Omega^2_3$ is the metric on the unit round three-sphere and $\ah$ is the scale factor. 

Quantum states of the universe are represented by wave functions $\Psi$ on the superspace spanned by the three-geometries and the field configurations on a spacelike surface $\Sigma$. Taking $\Sigma$ to be a surface of homogeneity in \eqref{homoiso}, useful coordinates on minisuperspace are the scale factor of the three-geometry which we denote by $b$ and the homogeneous value of the scalar field $\phi$ denoted by $\chi$. Histories of the form \eqref{homoiso} are specified by functions $(\ah(\lambda)$, $\phih(\lambda))$ that define curves in configuration space $(b(\lambda),\chi(\lambda))$ and vice versa.  Thus $\Psi=\Psi(b,\chi)$. When there is a need to be more compact we will write ${x^A} =(b,\chi), \  A=1,2$. Then $\Psi=\Psi(x^A)$. 

{\it Classical histories} of the form \eqref{homoiso} are specified by functions $\hat{a}(\lambda)$ and $\hat{\phi}(\lambda)$ that are real valued and satisfy the classical Einstein equations and the dynamical equation for the scalar field.
Classical histories are predicted in regions of superspace where the wave function is well approximated by a semiclassical (``WKB'') form
\be
\Psi(x^A) \approx A(x^A) \exp[i S(x^A)/\hbar] \ ,
\label{semiclassical-wf}
\ee
with $S$ varying rapidly compared to $A$ over the region. That is roughly when
\be 
\label{classcond}
|\nabla_A S| \gg |\nabla_A A/A|,  \quad \text{(Classicality Condition)}.
\ee
When these classicality conditions hold, a straightforward generalisation of WKB shows that a wave function \eqref{semiclassical-wf}  predicts an ensemble of classical histories of the universe that are the integral curves of $S$. That is, they are the solutions of  
\be
p_A = \nabla_A S\ ,
\label{momenta}
\ee
where $p_A$ are the momenta conjugate to $x^A$. The probability $p$ of the history that passes through $x^A$ is, to leading order in $\hbar$, given by 
\be
\label{probhist1}
p \propto |A(x^A)|^2 .
\ee
This is conserved along the classical histories as a consequence of the operator implementation of the Hamiltonian constraint --- the Wheeler-DeWitt equation. 

This semiclassical algorithm for classical prediction has been extensively used in quantum cosmology to extract predictions for cosmological observables from a wave function of the universe in domains where the classicality conditions hold. But histories need not be all quantum or all classical. Instead the classicality conditions \eqref{classcond} may hold in some regions of configuration space but not in other regions. 
Histories $(\ah(\lambda)$, $\phih(\lambda))$ do not end when the classicality conditions \eqref{classcond} break down. After all, histories are defined on the whole of the manifold $M={\bf R} \times S^3$. Instead when the classicality conditions fail deterministic classical evolution is replaced by quantum evolution. This allows for {\it quantum transitions} through the region of semiclassical breakdown that connect different parts of histories in the classical region of superspace \cite{Hartle:2015bna}. Tunneling through a barrier is a well known example of this.

In minisuperspace quantum cosmology, the quantum transition amplitude between two classical histories is specified by the propagator between an initial spatial hypersurface where $(b,\chi)=(b_{1},\chi_{1})$ and a final one with data $(b_{2},\chi_{2})$,
\be
\label{propagator}
T( b_{2},\chi_{2} | b_{1},\chi_{1}) \equiv \int_{(b_{1},\chi_{1})}^{(b_{2},\chi_{2})} \delta N \delta a \delta \phi   \exp{\{i {\cal S} [a,\phi]/\hbar\}} \,.
\ee
We show below that the transition probabilities derived from this stabilise as the boundary surfaces are moved further into the classical domain of the histories. Hence we obtain a transition matrix $T(b_{2},\chi_{2}| b_{1},\chi_{1})$ between classical histories, that in many ways is analogous to an S-matrix. The transition probabilities between specified classical histories are then proportional to
\be
\label{transprob}
p_{\rm trans}(b_{2},\chi_{2}|b_{1},\chi_{1}) \propto |T(b_{2},\chi_{2}|b_{1},\chi_{1})|^2 \,.
\ee
Below we calculate this propagator \eqref{propagator} in the semiclassical approximation in two different cosmological models.

%%%%%%%%%%%%%%%%%%%
\section{Quantum Transitions: from Inflation to Inflation} \label{inftoinf}
%%%%%%%%%%%%%%%%%%%

In this Section we consider a positive scalar potential and evaluate the propagator \eqref{propagator} in its saddle point approximation to compute quantum transitions connecting classical, inflationary histories across a de Sitter like throat or a classical singularity. In Section \ref{section:ekinf} we will return to the propagator \eqref{propagator} in models with more general potentials that allow for transitions between ekpyrotic contraction and inflationary expansion.

We first consider the semiclassical approximation to the propagator \eqref{propagator} interpolating between two identical inflationary histories on both ends. This amounts to finding (complex) `bounce' solutions of the Euclidean equations of motion of gravity coupled to a scalar field,
\begin{subequations}
\begin{align}
a'' +\frac{a \kappa^2}{3}\left(V(\phi)+ {\phi'}^2\right)=0  \ ,\\
\phi''+3\frac{a'}{a}\phi' -\frac{\partial V(\phi)}{\partial \phi}=0 \ ,\\
{a'}^2 -1 + \frac{\kappa^2 a^2}{3}\left(-\frac12 {\phi'}^2 +V(\phi)\right)=0 \ ,
\end{align}
\end{subequations}\label{Eeom}
where a prime denotes a derivative w.r.t. Euclidean time.

The symmetry of the problem means it is natural to consider saddle points that are symmetric around the bounce\footnote{This saddle point selection principle can be justified in the no-boundary quantum state. We return to this point below in Section \ref{nbwf-classens}.} This translates into the boundary condition that, at the point $\tau_s$ of symmetry, 
\begin{align}
a'(\tau_s)=0 \ , \qquad \phi'(\tau_s)=0 \ ,
\end{align}
where we are free to choose $\tau_s=0$. The complex value $\phi (\tau_s)$ of the scalar field at the surface of symmetry can be varied to obtain the required boundary values $(b,\chi)$ of the fields. The bounce value of the (complex) scale factor in turn is determined by the Hamiltonian constraint. Hence we have,
\begin{align}
\phi(\tau_s)=\phi_s e^{i \theta_s} \ , \qquad a(\tau_s)=\sqrt{\frac{3}{V(\phi(\tau_s))}} \ .\label{eqn:inita}
\end{align}
These constitute a sufficient set of boundary conditions to determine saddle point solutions to the propagator path integral. The bounce solutions can be viewed as solutions in the complex $\tau$-plane, with the bounce located at $\tau = \tau_s = 0$ and the boundaries where $(a,\phi)=(b,\chi)$ at some complex value $\pm \upsilon,$ with  $\upsilon=X+i t$. 

%%%%%%%%%%%%%%%%%
\begin{figure}[ht!]
	\centering
	\includegraphics[width=0.9\textwidth]{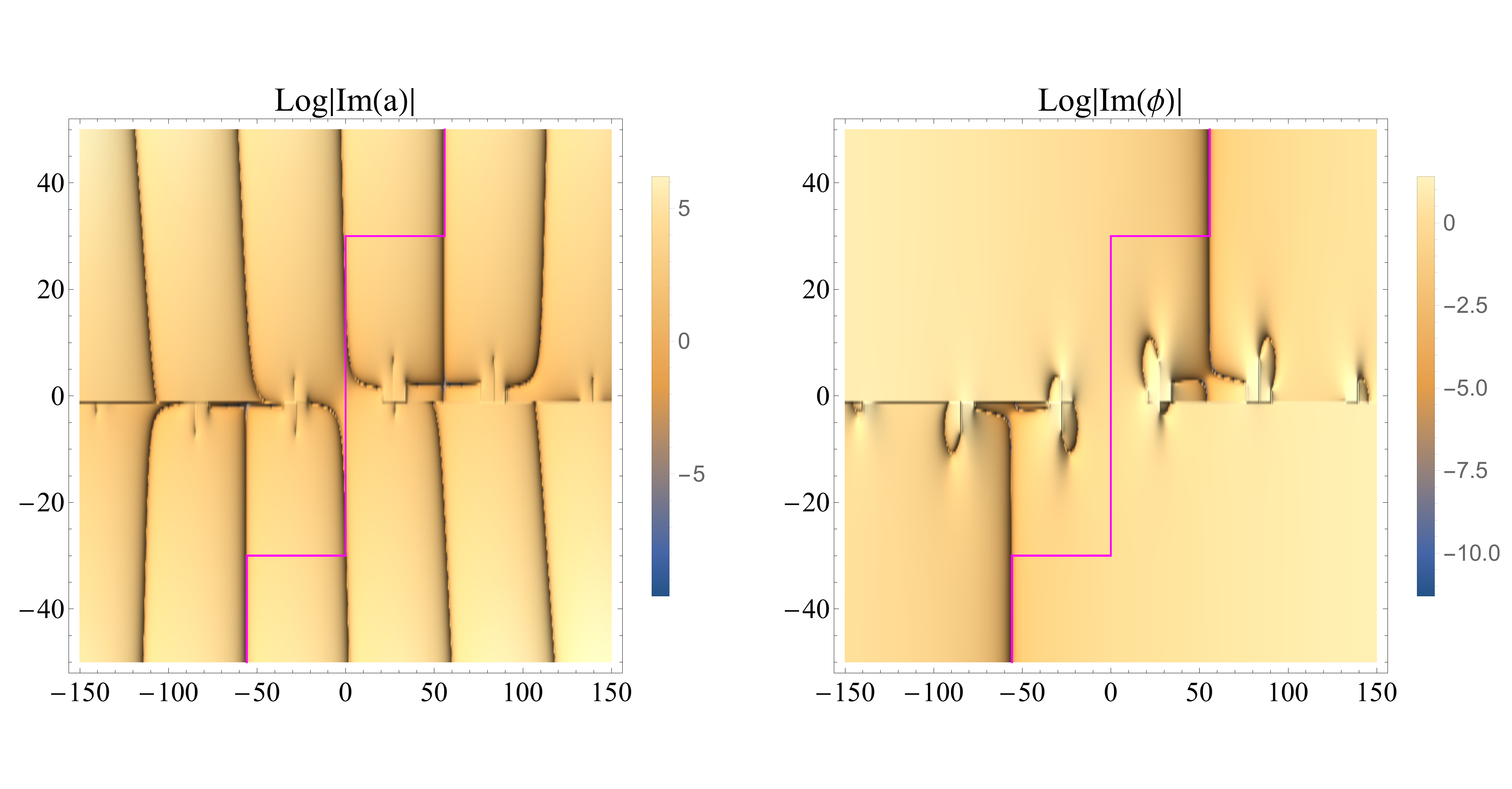}\\
	\includegraphics[width=0.9\textwidth]{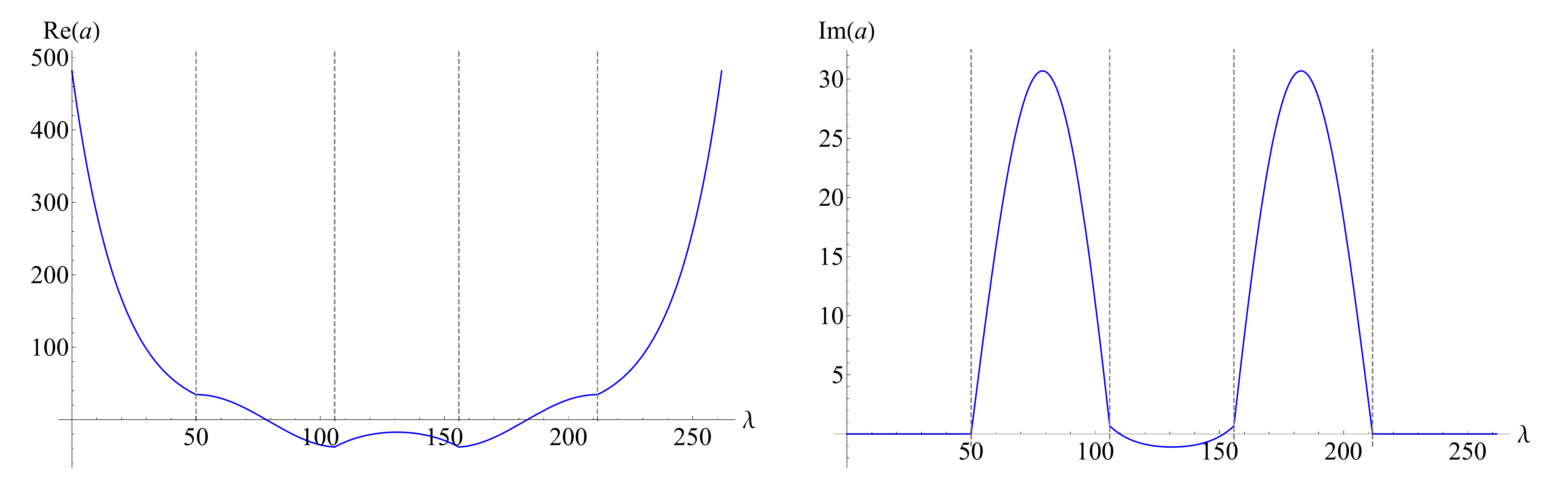}\\
	\includegraphics[width=0.9\textwidth]{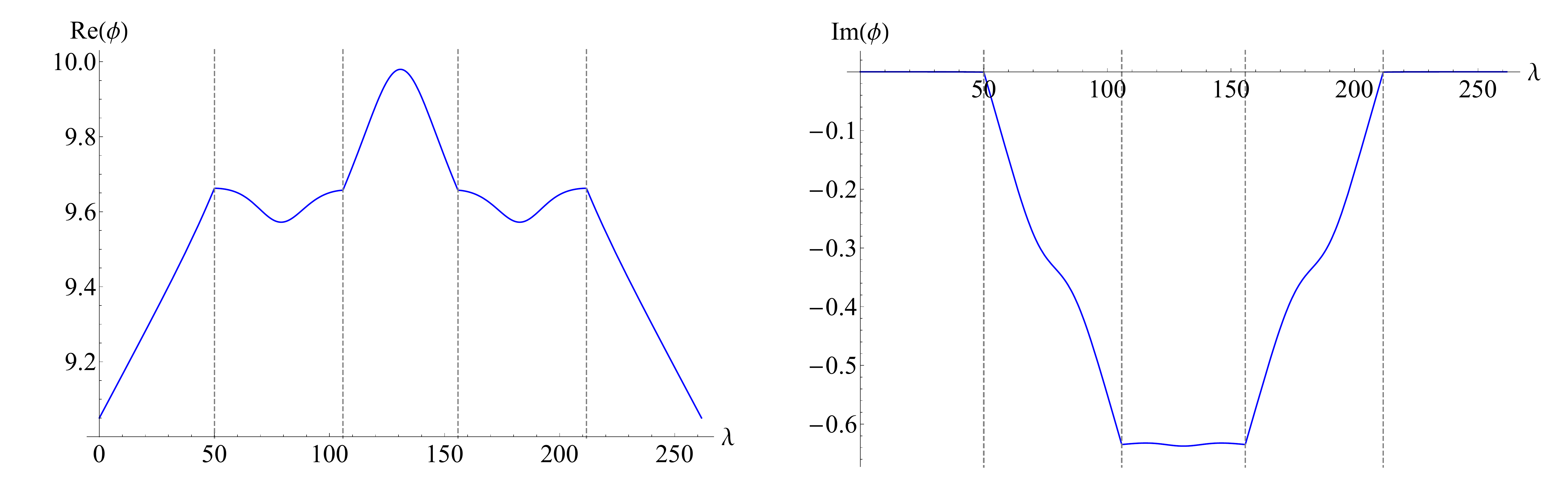}\\
	\caption{For $\phi_s=10, \theta_s= - 0.0637813$ we show on the left side $|\log(\textrm{Im}(a))|$ and on the right $|\log(\textrm{Im}(\phi))|$ in the complex $\tau$-plane. These plots show where the the scale factor and scalar field become real. The phase  $\theta_s$ is fine-tuned such that the lines of real values of $\phi$ and $a$ overlap on the first branch of $a$ to the right of the origin. The contour we chose is drawn in magenta, and the lower panels show the evolution of the fields along this contour as a function of $\lambda$, which is related to $\tau$ by $\tau=\int N d\lambda$. Here we took $\lambda=0$ in the bottom left corner and evaluated the fields from this point to the upper right corner where $\lambda=261.6$. For the present solution, we have $b_1=b_2=500,\, \chi_1=\chi_2=9.04196$ and at $(b_2,\chi_2)$ we have $a_{,\tau}=-  4.00006\times10^{-7}  - 26.1886 i,\,\phi_{,\tau}= 2.18152\times10^{-8}  +0.0115247 i.$}
	\label{fig:imvaluessymBounce}
\end{figure}
%%%%%%%%%%%%%%%%%%%%

To find the saddle points we must tune the three free parameters $\phi_s, \theta_s$ and $X$ such that the desired boundary values are reached. Fig. \ref{fig:imvaluessymBounce} shows an example, for a quadratic potential with $m=\sqrt{2} \cdot 10^{-2}$. The top panels show the logarithm of the absolute value of Im($\phi$) and Im($a$) in the complex $\tau$-plane for $\phi_s=10$ and $\theta_s= - 0.0637813$. Black lines correspond to large negative values of the logarithm, indicating where the fields become real. Loosely speaking, a classical history corresponds to having vertical black lines (vertical is the Lorentzian time direction) for $a$ and $\phi$ at the same location in the complex $\tau$-plane. It turns out that the scale factor has in general multiple lines parallel to the y-axis where its imaginary part becomes zero. This is because in a slowly changing potential, the solution for the scale factor is sinusoidal, $a \sim \sqrt{\frac{3}{V(\phi(\tau_s))}}\sin(\sqrt{\frac{V(\phi(\tau_s))}{3}}\tau).$ Then, by tuning the phase $\theta_s$ one can ensure the lines of real $a$ and real $\phi$ coincide. 
In fact, given there are multiple vertical lines where $a$ is real for a given $\phi_0$, one can find several symmetric bouncing saddle points connecting different classical solutions. In Fig. \ref{fig:imvaluessymBounce} we have taken $\theta$ such that the scalar field becomes real on the first branch in the upper right quadrant.

%%%%%%%%%%%%%%%%%%%%
\begin{figure}[ht!]
	\centering
	\includegraphics[width=0.9\textwidth]{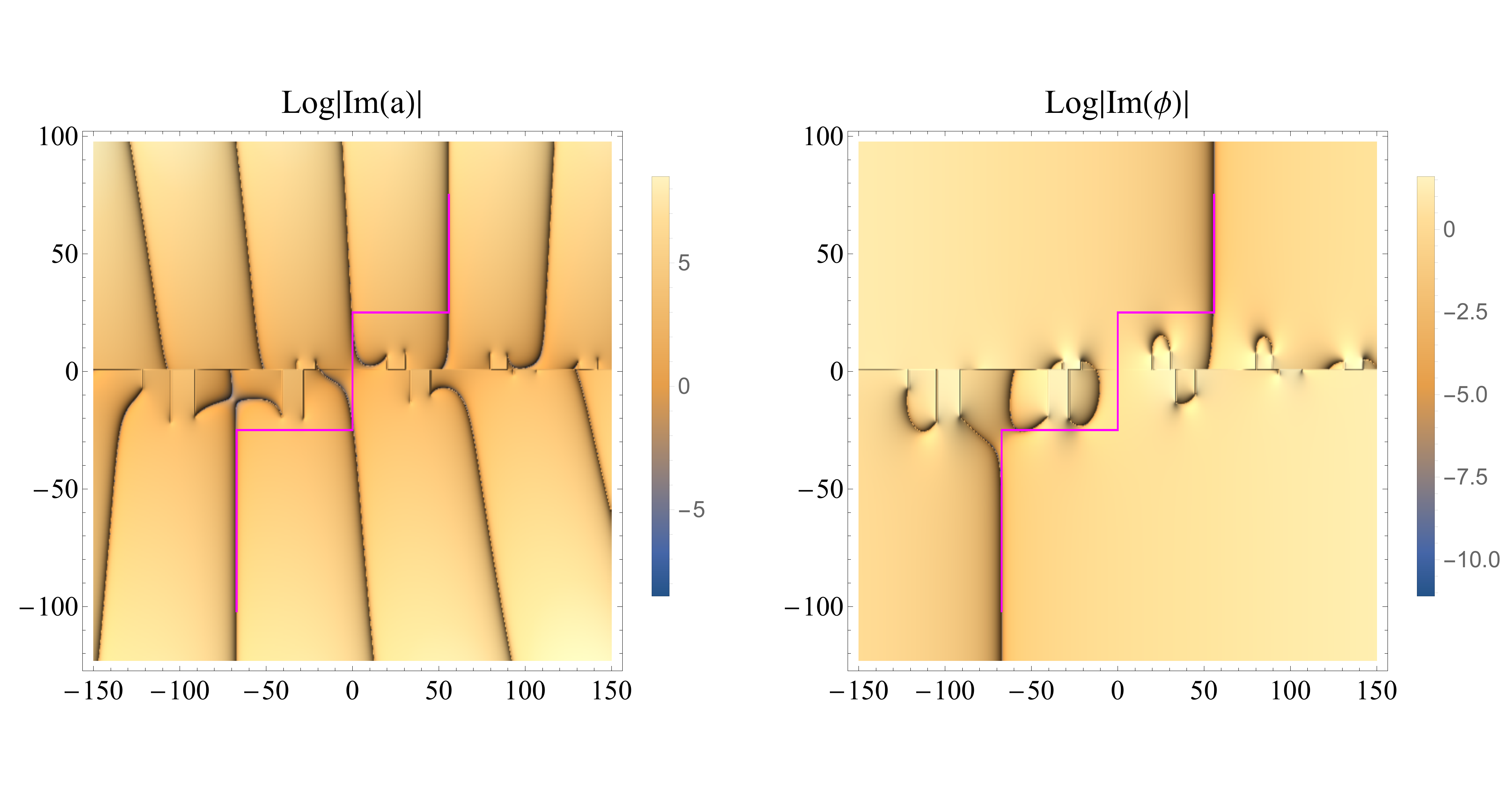}\\
	\includegraphics[width=0.9\textwidth]{infl-fields-sym-bounce-a.pdf}\\
	\includegraphics[width=0.9\textwidth]{infl-fields-sym-bounce-phi.pdf}\\
	\caption{An example of an asymmetric bounce with $(b_1=500, \chi_1=7)$ and $(b_2=500, \chi_2 = 9.04196)$. In the upper panel we show the behaviour of the imaginary parts of the fields in the complex $\tau$-plane, together with the contour we chose in magenta. In the lower panels we show the field values along this magenta path. This solution is obtained with the following derivatives imposed at the final boundary: $a_{,\tau}= 6.93303\times 10^{-6}   - 26.1886 i,\,\phi_{,\tau}= - 3.78052\times10^{-7}  +0.0115265 i.$}
	\label{fig:imvaluesasymBounce}
\end{figure}
%%%%%%%%%%%%%%%%%%%%

The action of the complex bouncing saddle points determines the quantum transition amplitude between the two classical histories at the endpoints. We have chosen to integrate the action along a contour that is not only symmetric, as required by the NBWF, but which also provides the dominant contribution to the quantum transition. In appendix \ref{contour} we discuss our choice of contour in more detail. We also discuss approximate analytic solutions in appendix \ref{analytic}.

The contour we selected is shown in Fig.  \ref{fig:imvaluessymBounce} in magenta in the upper panel. The evolution of the fields along this contour is shown in the lower panels of the figure. The fields indeed become real along the last vertical leg out to the end points, where the saddle points coincide with a classical history. The upper panel shows there are singularities in the complex $\tau$-plane, especially along the real $\tau$ axis. Had we chosen a contour encircling one of these singularities, we would have obtained either a different solution or no solution at all.

%%%%%%%%%%%%%%%%%%%%
\begin{figure}[ht!]
	\centering
	\includegraphics[width=0.6\textwidth]{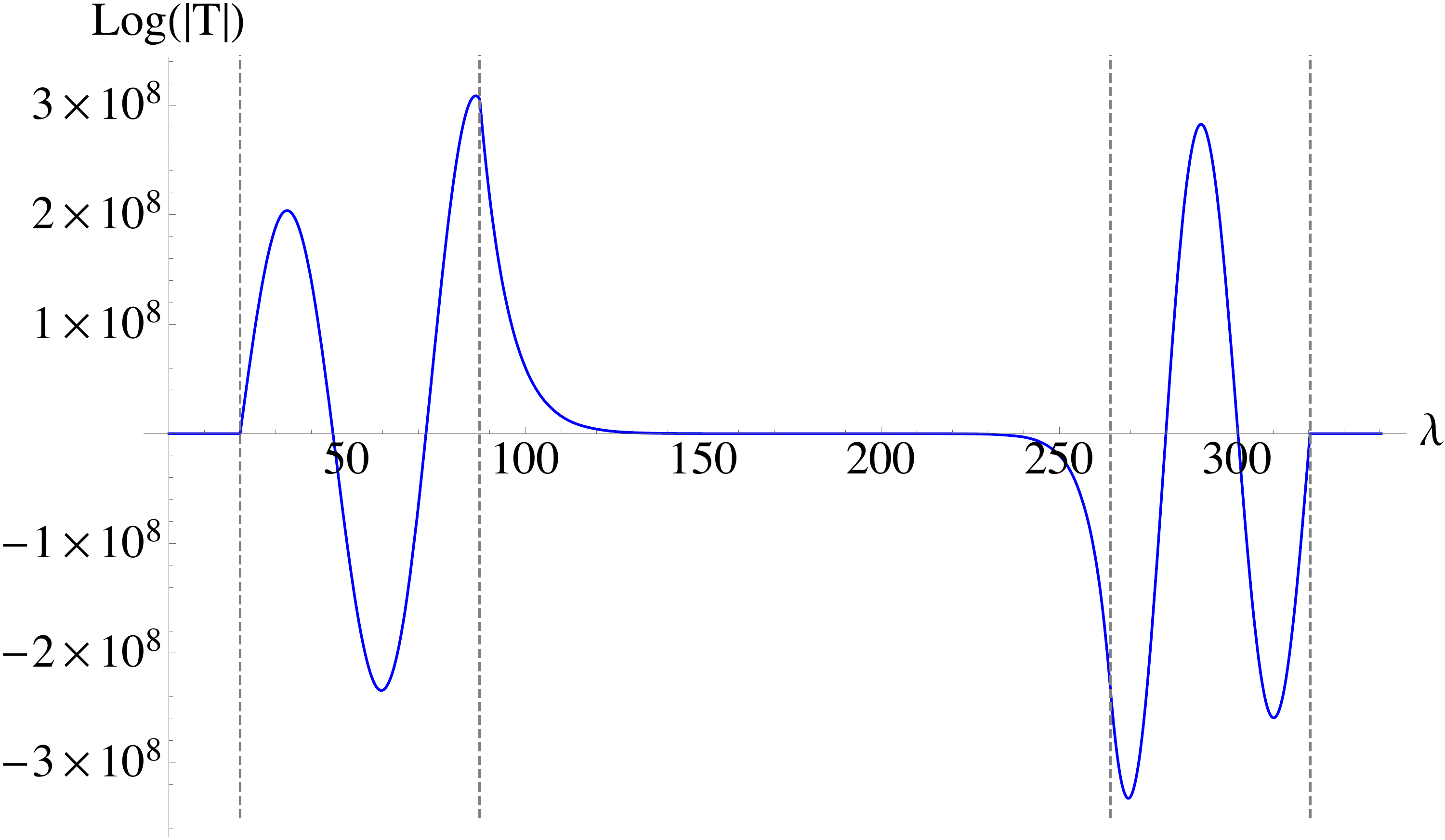}
	\caption{The real part of the interpolating saddle point action along the magenta contour shown in Fig. \ref{fig:imvaluesasymBounce}. This tends to a constant near the boundaries of the instanton, indicating that the quantum transition probabilities between given classical histories rapidly stabilise in the classical domain away from the bounce.}
	\label{fig:realaction}
\end{figure}
%%%%%%%%%%%%%%%%%%%%

Evidently the symmetric bounce solutions can equally well be obtained by integrating the equations of motion from one of the boundaries, instead of starting at the point of symmetry. This proves to be a more useful setup to find interpolating saddle points between different classical histories on both ends. Fig. \ref{fig:imvaluesasymBounce} shows an example of such an asymmetric bounce, connecting two inflationary histories with a different number of efolds. The contour we selected to compute this asymmetric transition is the one which smoothly changes into the original symmetric contour, without crossing any singularities, when the data on both boundaries are taken to be equal again. 

At this point one may wonder whether the quantum transition probabilities \eqref{transprob} depend on the boundary value that is taken for the scale factor in the calculation of the propagator \eqref{propagator}. Clearly this should not be the case as long as the classicality conditions hold on the boundary, since classical evolution preserves the real part of the Euclidean action.

%%%%%%%%%%%%%%%%%%%%
\begin{figure}[ht!]
	\centering
	\begin{minipage}{0.48\textwidth}
	\includegraphics[width=0.9\textwidth]{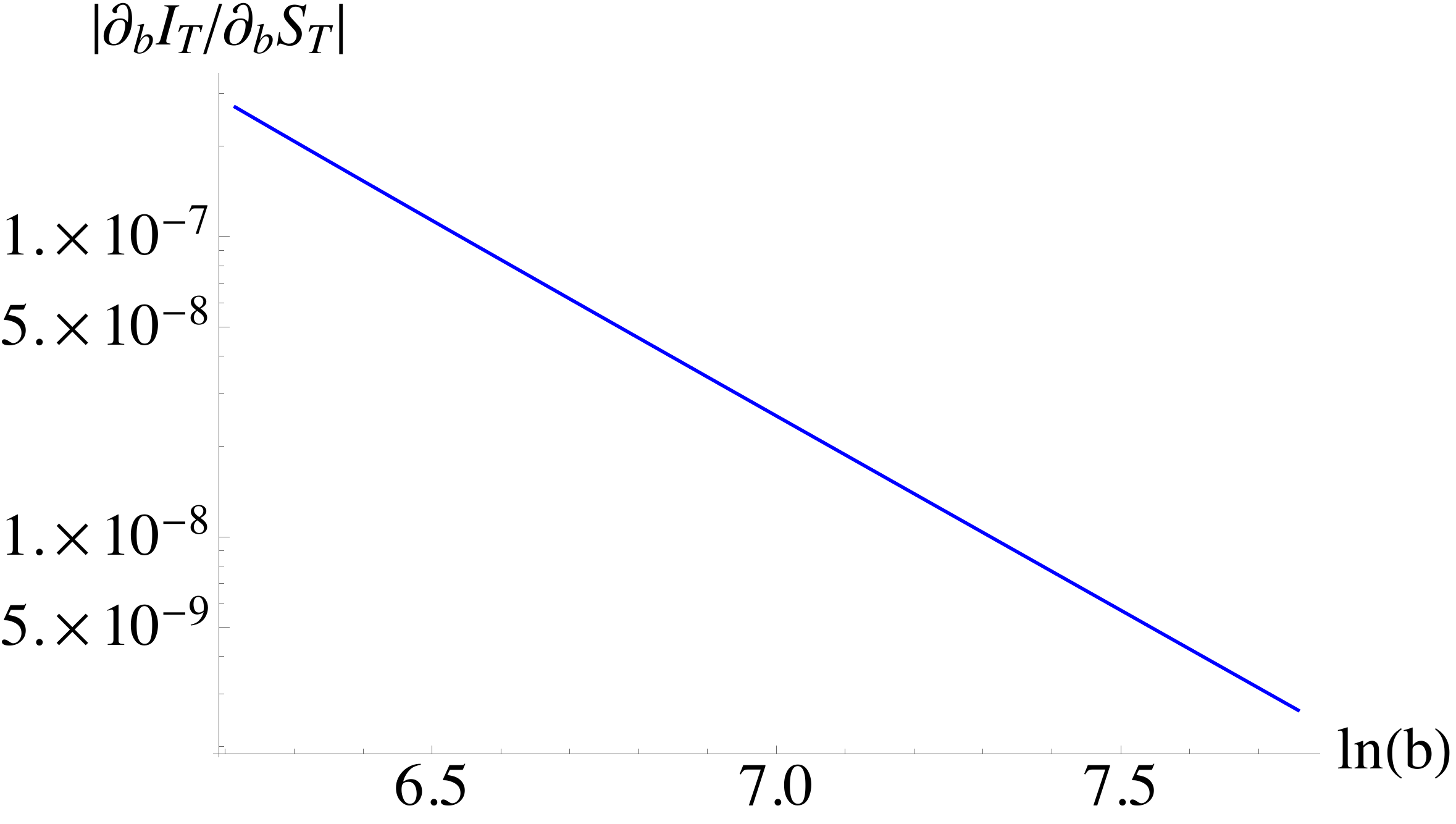}
	\end{minipage}
	\begin{minipage}{0.48\textwidth}
	\includegraphics[width=0.9\textwidth]{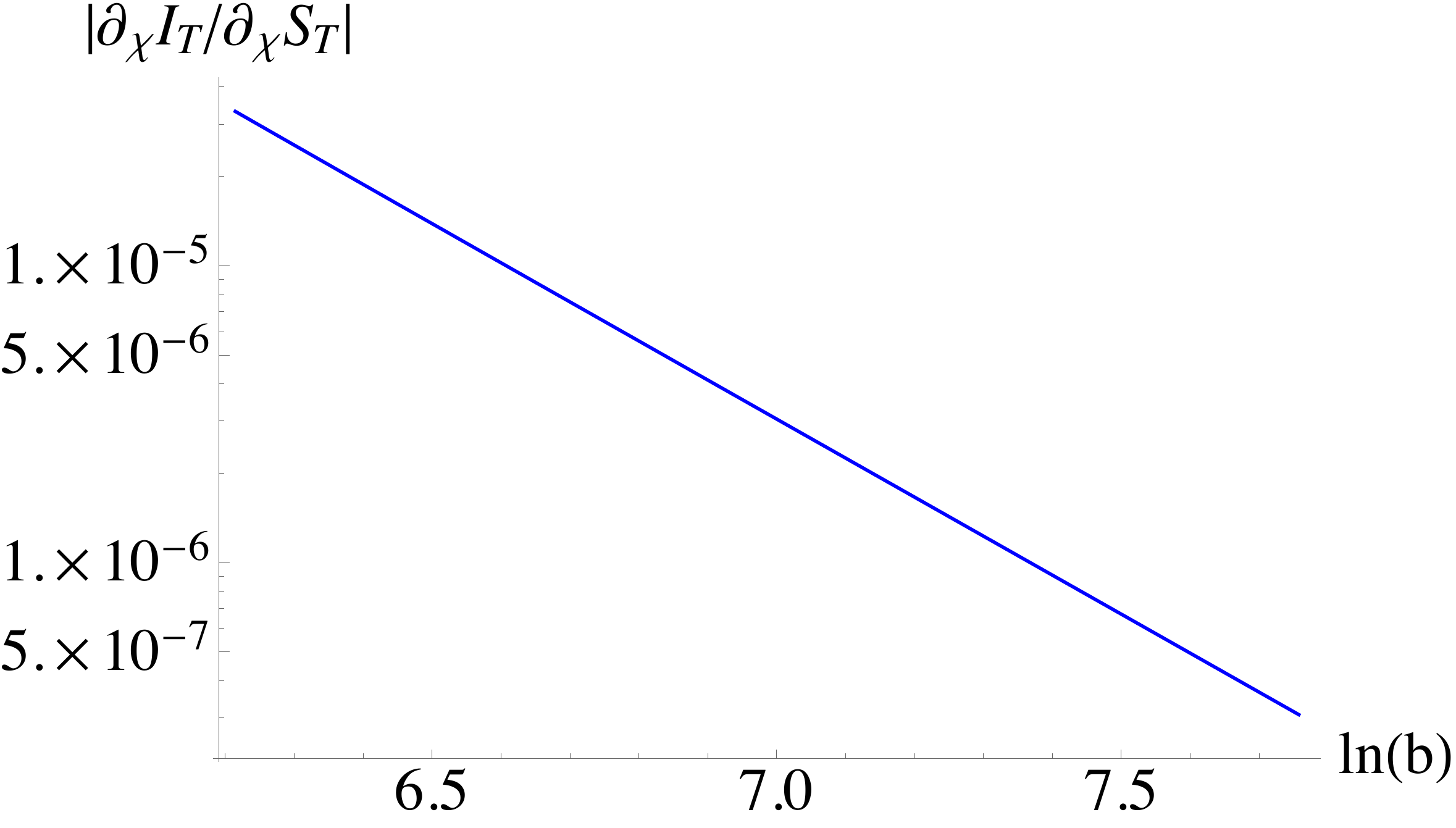}
	\end{minipage}
	\caption{Classicality conditions derived from a saddle point describing a quantum bounce, assuming a classical incoming history, as a function of $\ln(b)$. The smallness of these ratios shows that the transition probability stabilises along the outgoing classical history.}
	\label{fig:cc}
\end{figure}
%%%%%%%%%%%%%%%%%%%%

It is therefore a useful consistency check of our method to verify whether the resulting transition probabilities stabilise if we take the boundary surfaces to larger scale factor, deeper into the classical domain of the histories.  A first indication that this will indeed be the case for our solutions is provided in Fig. \ref{fig:realaction} which shows, for the solution plotted in Fig. \ref{fig:imvaluesasymBounce}, that the real part of the Euclidean action of the interpolating saddle point tends to a constant near both boundaries. A more precise assessment is given in Fig. \ref{fig:cc} where the WKB ratio $\nabla_A I_T/\nabla_A S_T$ is plotted as a function of $b.$ We defined here $I_T$ and $S_T$ as respectively the real and imaginary part of the Euclidean action. The derivatives are estimated from taking finite differences, obtained by calculating successive interpolating instantons matching onto a classical history $\left( b(\lambda), \chi(\lambda)\right)$, as well as slightly displaced instantons $\left( b+ \delta b, \chi \right)$ and $\left( b, \chi + \delta \chi\right).$ The fact that the WKB conditions are small means the real part of the action is conserved. Thus the transition probability is independent of the slice at which the interpolating instanton is matched onto a given classical history. This also means that to compute asymmetric transitions we can fix $b$ to a convenient value, and let the boundary values of the scalar field vary.

%%%%%%%%%%%%%%%%%%%%
\begin{figure}[ht!]
	\centering
	\includegraphics[width=0.75\textwidth]{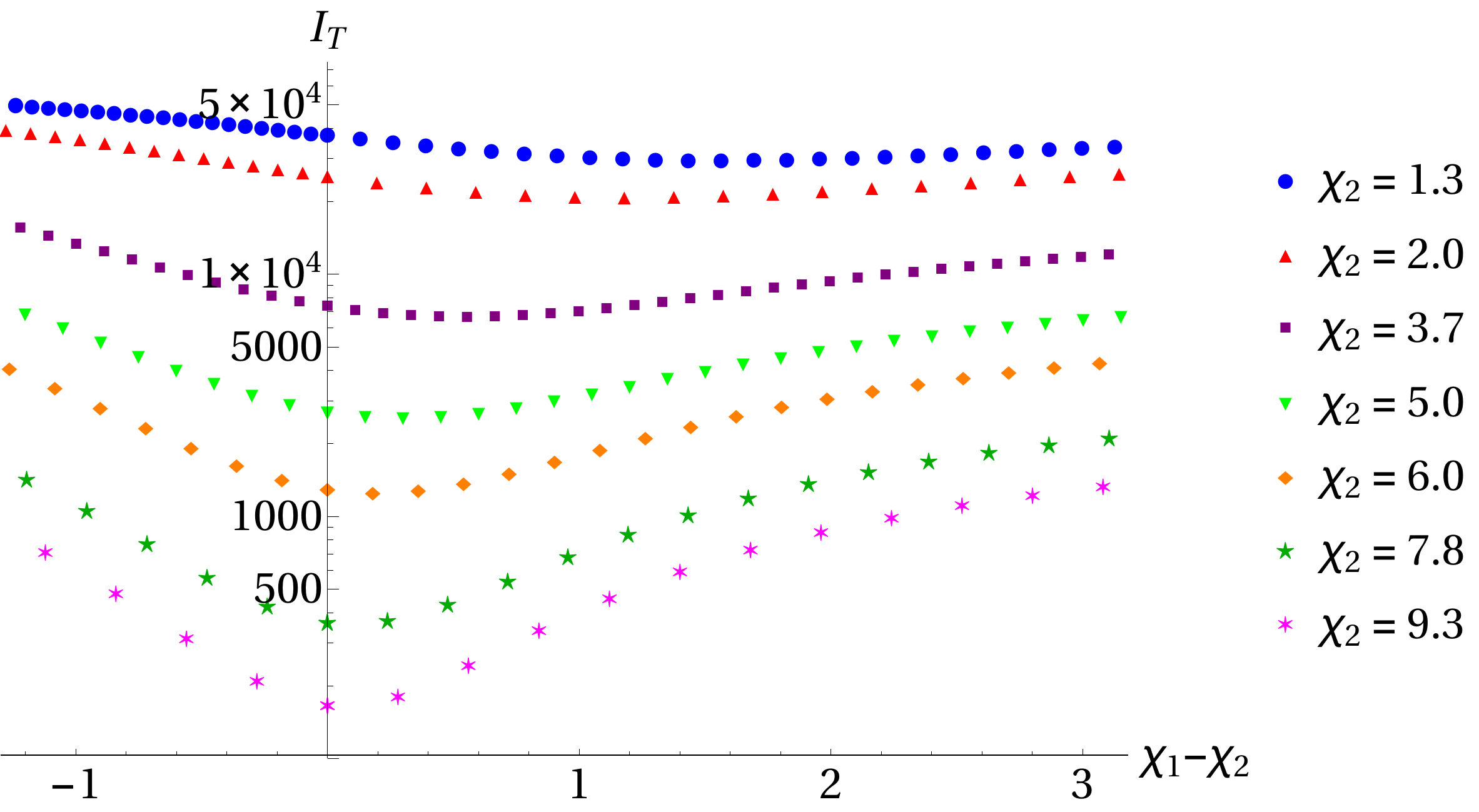}
	\caption{Logplot of the real part of the transition actions $I_R$ as a function of the difference $\chi_1-\chi_2$ for seven different initial values $\chi_2$ shown in the legend. On both sides of the transition we fixed $b=500$, well into the classical regime.}
	\label{fig:actionschi1}
\end{figure}
%%%%%%%%%%%%%%%%%%%%

\begin{figure}[ht!]
	\centering
	\includegraphics[width=0.75\textwidth]{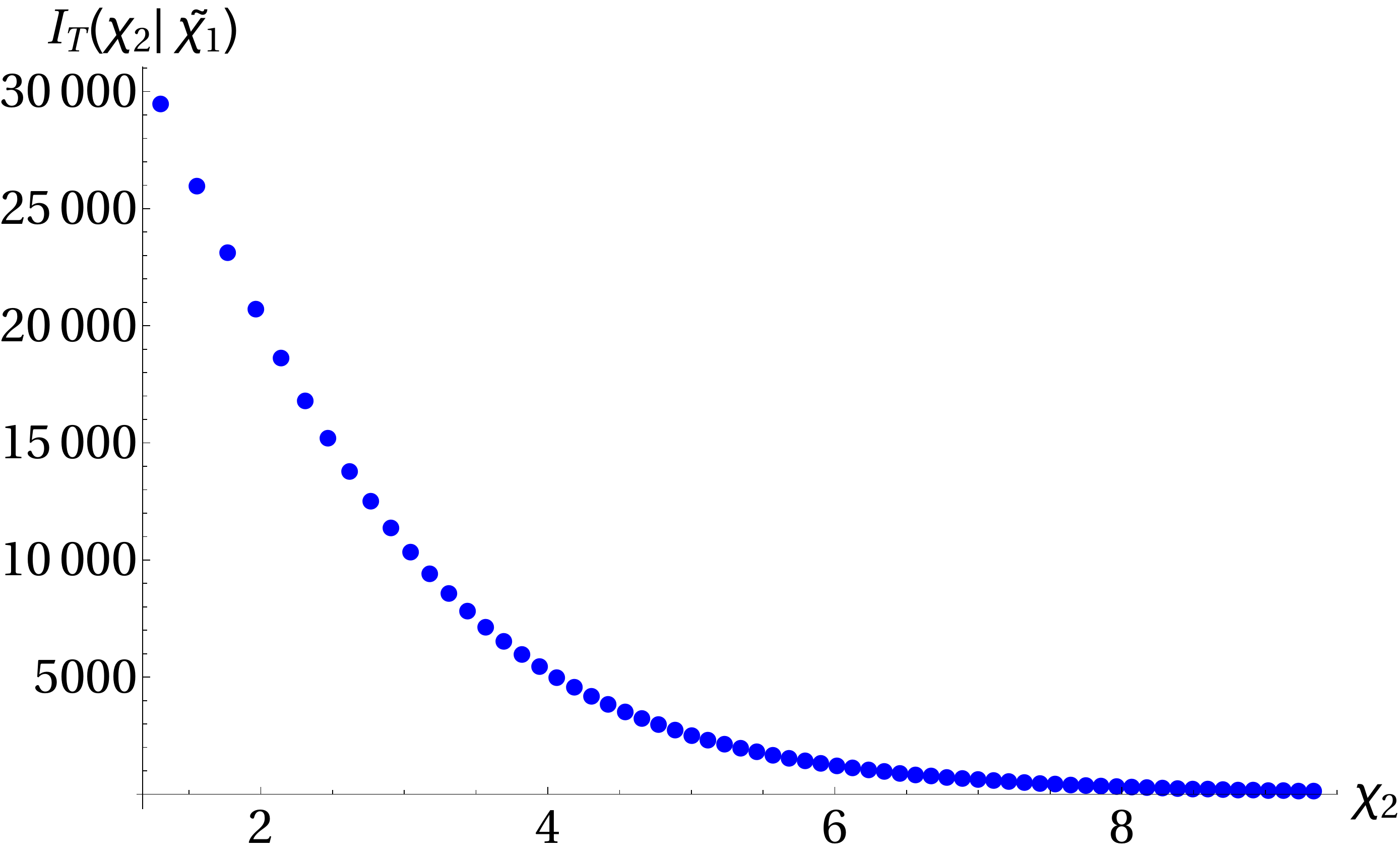}\\
	\caption{The real part of the transition action, as a function of $\chi_2$, for the most probable history on the other side of the bounce. The values $\tilde{\chi}_1$ thus correspond for each $\chi_2$ to the minimum of the curves in Fig. \ref{fig:actionschi1}. This shows that transitions are more likely for larger $\chi_2$.}
	\label{fig:transitionifochi2}
\end{figure}

Our results for the semiclassical quantum transitions between inflationary histories in a quadratic potential are summarized in Figures \ref{fig:actionschi1} and \ref{fig:transitionifochi2}. Shown in Fig. \ref{fig:actionschi1} are the real part of the saddle point actions interpolating to different final values of $\chi_2= 9.3, 7.8, 6.0, 5.0, 3.7, 2.0, 1.3 $, as a function of the difference $\chi_2-\chi_1$ with fixed $b=500$ on both sides of the bounce. For large initial scalar field values $\chi_2$ the most probable transition is the symmetric one. If we decrease $\chi_2$ the instanton actions increase, giving a lower overall probability for a transition to occur. This behaviour is similar to that of probability distributions resulting from the tunneling wave function in cosmology \cite{Vilenkin:1983xq}, which one might have expected since the transitions we compute are not unlike tunneling events. This is illustrated in Fig. \ref{fig:transitionifochi2} where we integrated over all initial values $\chi_1$ to obtain the total probability to transition as a function of $\chi_2$.  Finally we note that the minima shift slightly towards larger values of $\chi_1-\chi_2$, implying that transitions to universes with a slightly longer period of inflation are preferred.

%%%%%%%%%%%%%%%%%%%%
%%%%%%%%%%%%%%%%%%%%

Transition probabilities of this kind can be used to compute probabilities for entire four-dimensional histories in any quantum state that predicts classical inflationary patches. We now consider a specific example.

%%%%%%%%%%%%%%%%%
\section{Bouncing Inflationary Histories}
\label{nbwf-classens}
%%%%%%%%%%%%%%%%%

In this Section we combine our results for quantum transitions between inflationary universes with the NBWF, which provides a quantum mechanical framework of inflationary cosmology \cite{Hartle2007,Hartle2008,Lehners:2015sia}. This yields probabilities for an ensemble of complete inflationary histories exhibiting a quantum transition that connects two classical inflationary regimes on either side.

%%%%%%%%%%%%%%%%%
\subsection{Classical Predictions of the No-Boundary State}
\label{nbwf-classens}
%%%%%%%%%%%%%%%%%

The semiclassical NBWF specifies a measure on the minisuperspace configuration space $(b,\chi)$ in which individual histories are weighted by $\exp(-I/\hbar)$, where $I[g,\phi]$ is the Euclidean action of a regular and compact `saddle point solution' that matches $(b,\chi)$ on its only boundary $\Sigma$. 

The Euclidean minisuperspace action $I[a,\phi]$ of gravity coupled to a single scalar field reads\footnote{We work in $\hbar=c=G=1$ Planck units from now on.}
\begin{align} \label{curvact}
I[a,\phi]=\frac{6 \pi^2}{\kappa^2}\int d\tau \left(-a {a'}^2-a+\frac{\kappa^2 a^3}{3}\left(\frac12 {\phi'}^2
+V(\phi)\right)\right) \,,
\end{align} 
and the resulting equations of motion were given in \eqref{Eeom}.

The regularity of the saddle points together with the conditions that $a=b$ and $\phi=\chi$ on $\Sigma$ mean the interior solutions are generally complex. Like the bounce solutions above, the saddle points defining the semiclassical NBWF can be thought of as solutions in terms of a complex time coordinate $\tau$\footnote{We refer the reader to \cite{Hartle2007,Hartle2008,Lehners:2015sia} for a detailed description and analysis of the semiclassical NBWF.}.
The semiclassical NBWF thus takes the form 
\begin{equation}
\Psi(b,\chi) \approx  \exp[-I(b,\chi)/\hbar] = \exp\{[-I_R(b,\chi) +i S(b,\chi)]/\hbar\} ,
\label{semiclass}
\end{equation}
where $I_R(b,\chi)$ and $-S(b,\chi)$ are the real and imaginary parts of the Euclidean action evaluated at the saddle point.
In regions of superspace where $S$ varies rapidly compared to $I_R$, the NBWF predicts an ensemble of classical histories, one through each point $(b,\chi)$ in the region where the classicality conditions hold, with relative probabilities determined by $\exp[-2I_R(b,\chi)/\hbar]$.

%%%%%%%%%%%%%%%%%%%%
\begin{figure}[ht!]
	\centering
	\includegraphics[width=0.48\textwidth]{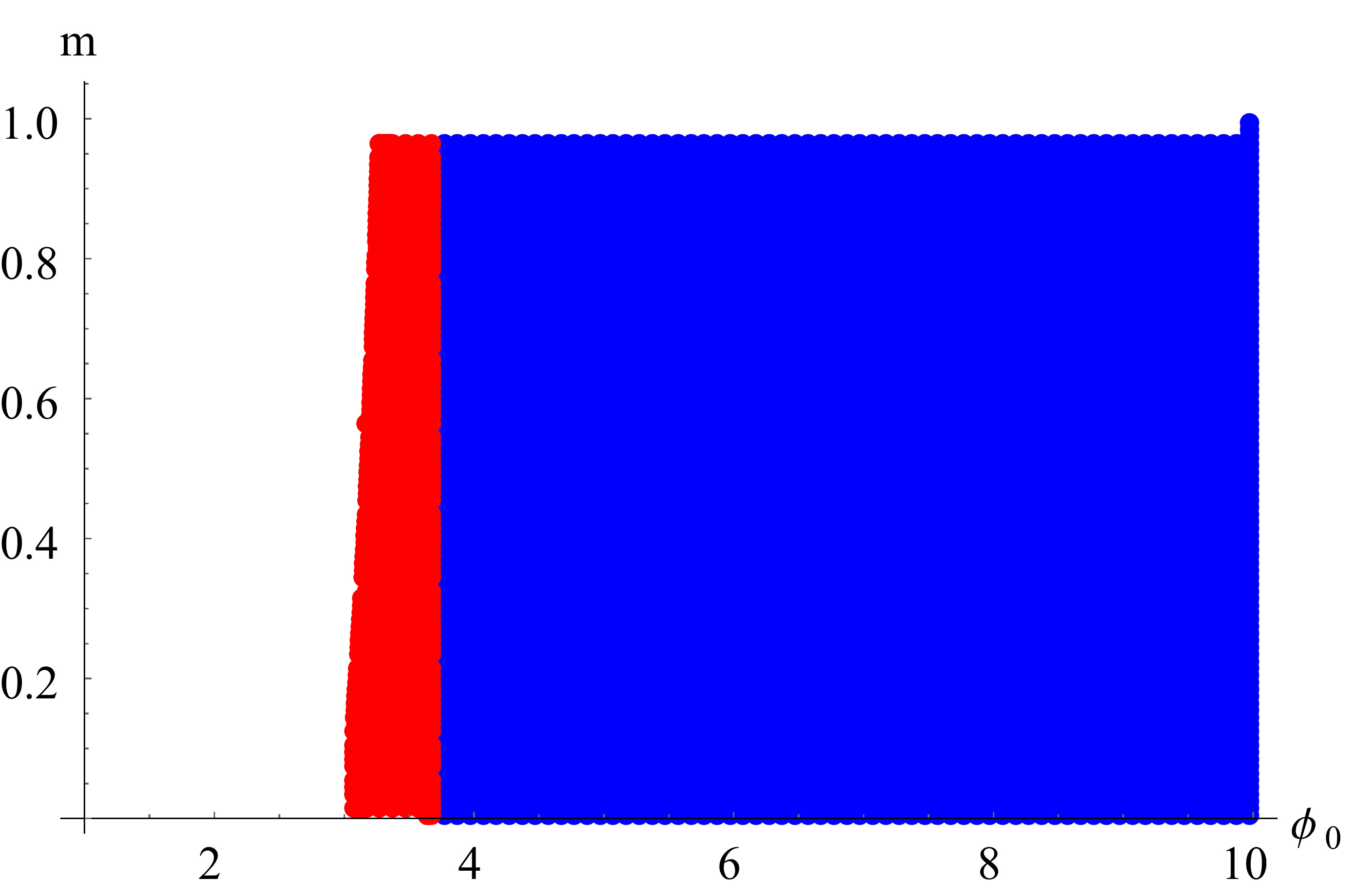}
	\includegraphics[width=0.48\textwidth]{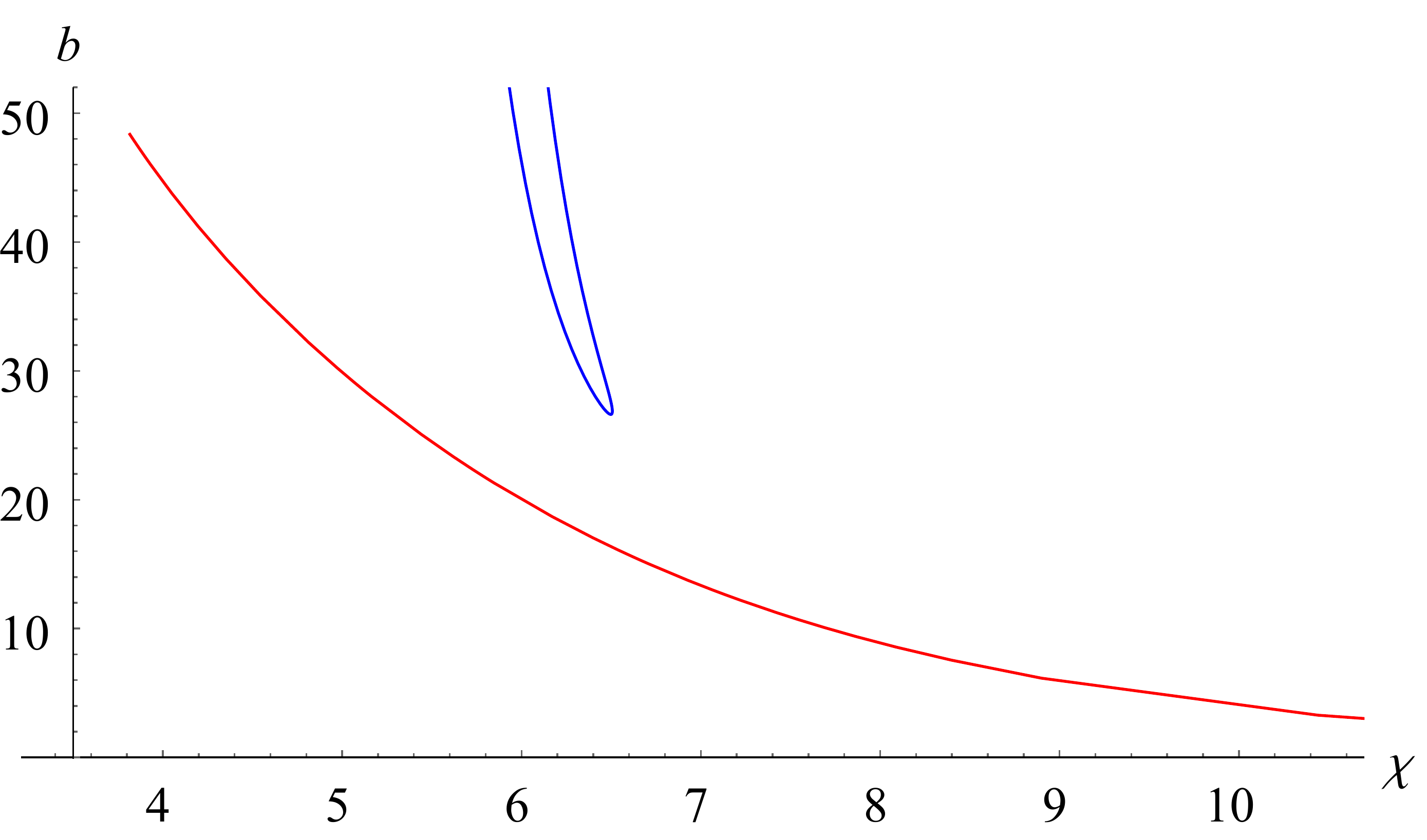}\\
	\caption{{\it Left panel}: The coloured regions are regions of parameter space $(m,\phi_0)$ that admit classical, inflating histories in the no-boundary state. The value $\phi_0$ corresponds roughly to the initial value of the scalar at the start of inflation and $m$ is the mass of the field. Histories in the region in red have an initial singularity when extrapolated classically backwards in time, whereas the classical histories in the blue region have a regular bounce in the past. {\it Right panel:} Two examples of classically extrapolated histories for $m=\frac{\sqrt{2}}{100}$; an initially singular history (red) and a bouncing history (blue).} \label{classhist}
\end{figure}
%%%%%%%%%%%%%%%%%%%%

The regularity condition on the saddle points selects a one parameter subset of homogeneous isotropic histories. These can either be labeled by the value of $\chi$ at some fiducial value of $b$,  or alternatively by their value of $\p0 \equiv \vert \phi(\tau=0) \vert$, the absolute value of $\phi$ at the south pole (SP) of the saddle point, where $a=0$, to which the classical history corresponds \cite{Hartle2008}. If $V(\phi)$ is everywhere positive then the classical NBWF ensemble has the remarkable property that {\it all} histories have an early period of scalar field driven inflation \cite{Lyons:1992ua,Hartle2008}. For instance, for a quadratic potential the solutions take the approximate form
\begin{subequations}
\label{slowroll}
\begin{align}
\chi(t) & = \p0 -mt/3 , \\
b(t)&= \frac{1}{2m\p0}\exp\left(m\p0 t -\frac{1}{6}m^2 t^2\right) \ ,
\label{slwroll}
\end{align}
\end{subequations}
which is just a subset of the usual family of inflationary slow roll solutions.

The number of efolds $N$ is $N \sim \p0^2$. Fig \ref{classhist}(b) shows two examples of classical histories in the NBWF ensemble for this minisuperspace model. The trajectories shown in this figure were obtained by naively extrapolating the histories \eqref{slowroll} both to the future and into the past of the slow roll era using the classical equations. For values $\p0 \gg {\cal O}(1)$ the classically extrapolated histories bounce at a minimum radius $b_m \approx  (m \p0)^{-1}$, never reaching a singularity \cite{Hartle2008}. By contrast for $ 3.1 \lesssim \p0  \lesssim 3.7 $ the classical histories begin with a singularity at infinite scalar field and zero scale factor and expand forever.

However the classicality conditions \eqref{classcond} hold neither near the minimum radius nor near the singularity \cite{Hartle2007,Hartle2008}. The classical extrapolation of the individual histories is thus not justified and must really be replaced by quantum evolution.

Where could quantum evolution lead to? At this point we must remember the NBWF is real. This means that for every saddle point of the form \eqref{semiclass} there is a complex conjugate saddle point with $S$ replaced by $-S$. Reversing the sign of $S$ amounts to reversing the direction of time in the corresponding classical history [cf. \eqref{momenta}]\footnote{See \cite{Hartle:2015bna} for a more extensive discussion of this point.}. For every classical history in the first ensemble, therefore, its time reversed is in the second. The NBWF thus predicts two identical copies of a set of inflationary histories, with one set interpreted as expanding histories and the other set as contracting histories. Both sets are not connected classically in the no-boundary state, because the classicality conditions fail. But as we now show, we can use our results for the propagator \eqref{propagator} above to `connect' a classical history in one ensemble by a quantum transition to a classical history in the other ensemble to make a complete history of the universe.

%%%%%%%%%%%%%%%%%%%%
\subsection{Probabilities for Bouncing No-Boundary Histories}\label{sec:prob}
%%%%%%%%%%%%%%%%%%%%

Probabilities for entire histories of our universe that are in the classical NBWF ensemble in both asymptotic regions and that exhibit a quantum transition connecting both asymptotic regions take the form \cite{Hartle:2015bna},
\be
\label{probhist2}
p(\p0'',\p0') = \pnb(\p0'')p_{\rm trans}(\p0'',\p0')\pnb(\p0') \ .
\ee 
The labels $\p0'$ and $\p0''$ refer to the absolute values of the scalar field at the SP of the saddle points that correspond to the classical histories in both asymptotic regions, in the same way we had $\chi_1$ and $\chi_2$ in the previous sections, and $\ptra$ is the transition probability between those histories discussed in Section \ref{inftoinf}.

The expression \eqref{probhist2} is symmetric under interchanging $\p0'$ and $\p0''$, or, put differently, under interchanging expanding and contracting. The ensemble of bouncing histories in the NBWF is therefore time symmetric by construction, respecting the quantum mechanical symmetry of the wave function. The individual histories are not symmetric, however, because the asymptotic behaviours of the expanding and contracting classical ends are different for different $\p0$. But for every history in the ensemble the time-reversed is also in the ensemble, with equal probability. 

\begin{figure}[ht!]
	\centering
	\includegraphics[width=0.49\textwidth]{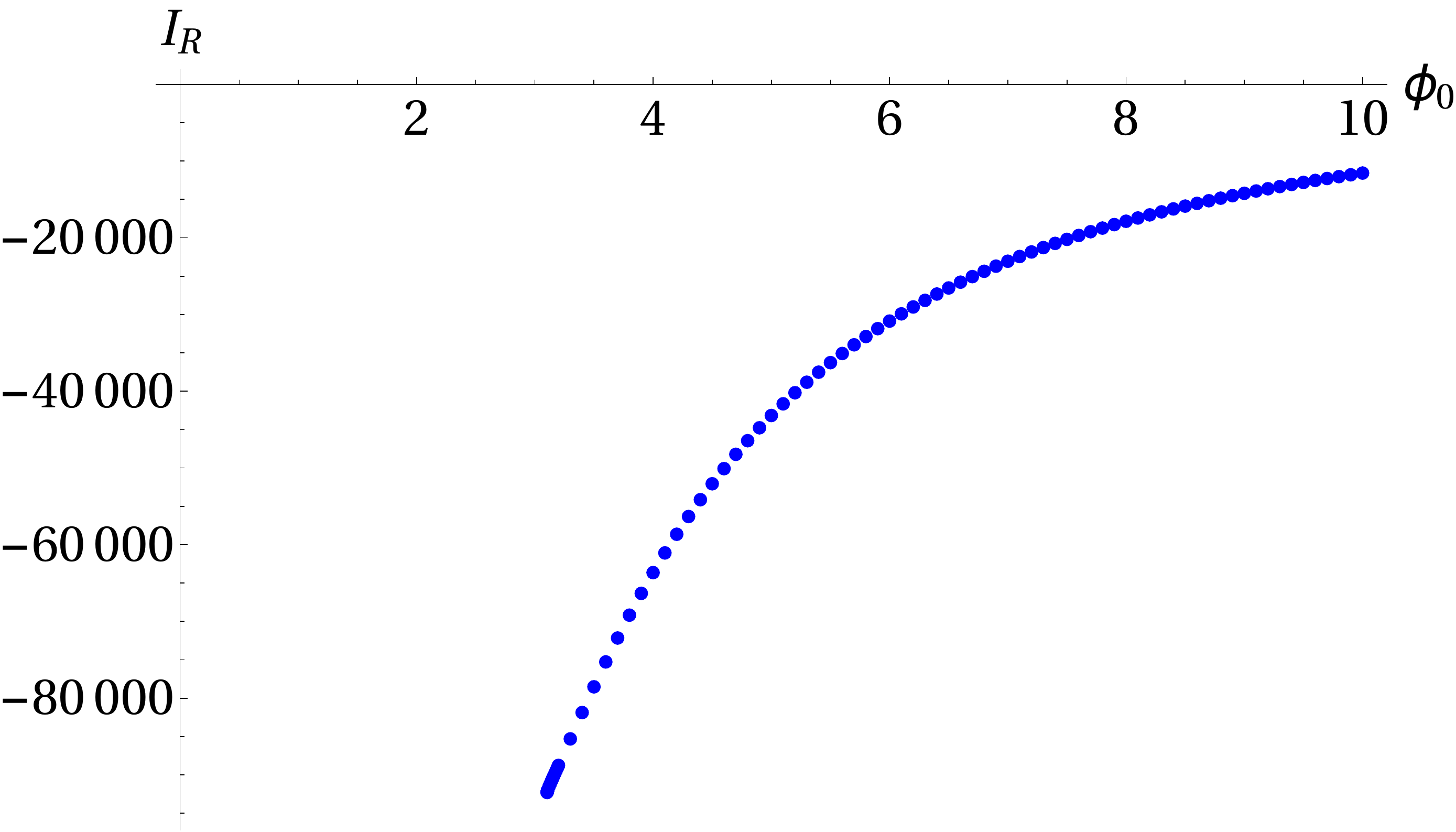}
	\includegraphics[width=0.44\textwidth]{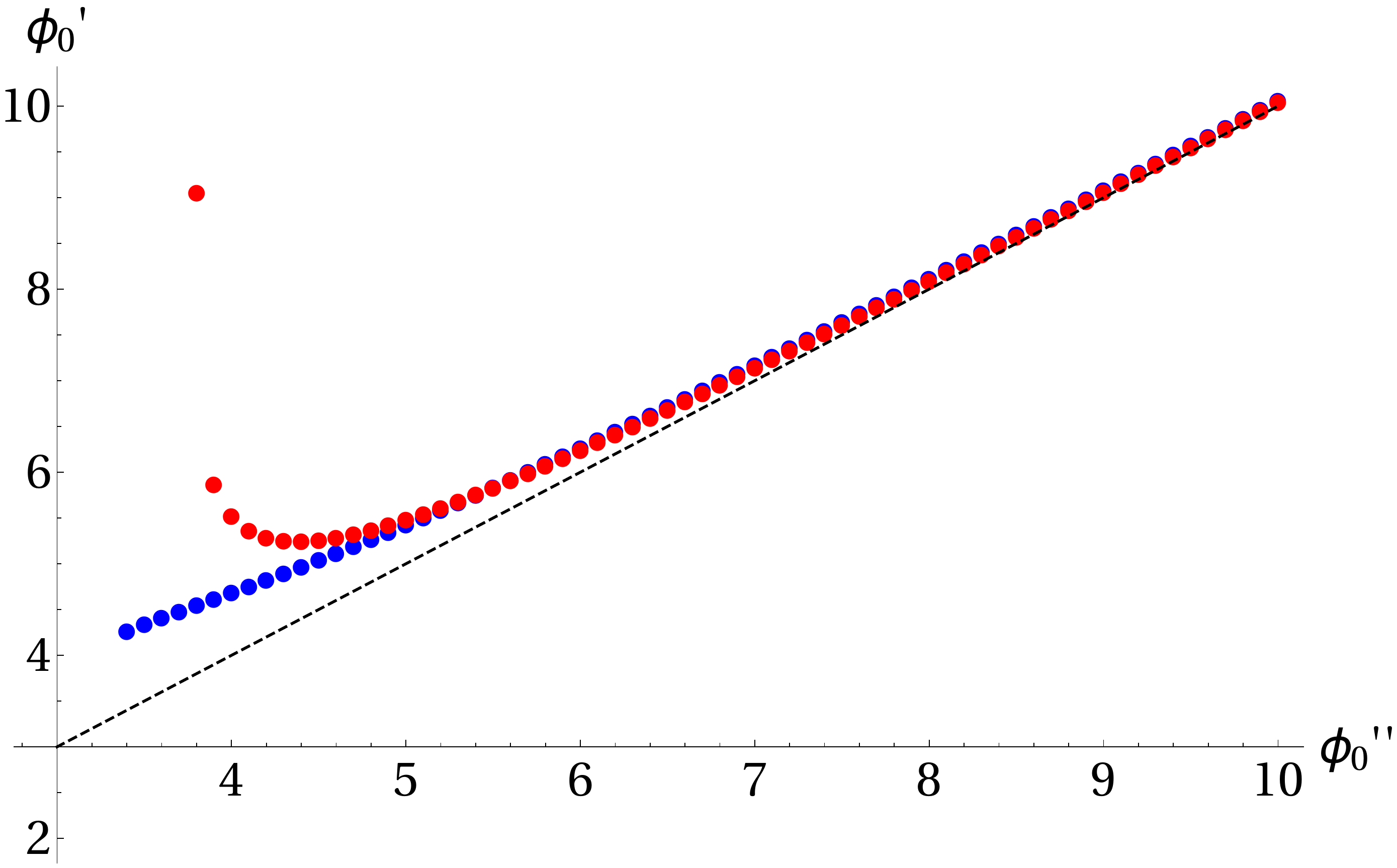}\\
	\caption{{\it Left panel:} The relative probabilities $p_{NB}(\p0)$ of classical inflationary histories in the NBWF as a function of $\phi_0$, the absolute value of the scalar field at the SP of the corresponding no-boundary saddle point, for a quadratic potential with $m^2=2.10^{-4}$.\\ {\it Right panel:} The most probable history (labeled by $\phi_0'$) on the other side of the quantum bounce as a function of the history on our side (labeled by $\phi_0''$) is given in blue whereas the $\phi_0'$ resulting from the classical extrapolation of our side backwards in time is given in red. The black dotted line indicates where symmetric bounces would be. The classical extrapolation produces a classical bounce for large $\p0''$ but becomes singular for small values of $\p0''$.} \label{fig:noboundaryWeight}
\end{figure}

The NBWF probability $p_{NB}(\p0)$ of an asymptotically classical inflationary region labeled by $\p0$, regardless its past, is approximately given by
\be
\label{nbprob}
p_{NB}(\p0) \propto \exp[-2I_R(\p0)] \approx \exp[+3\pi/V(\p0)] \ .
\ee
It is well known that the NBWF strongly favors histories with a low amount of slow roll inflation. This is illustrated in 
Fig.\ \ref{fig:noboundaryWeight}(a) for a quadratic scalar potential. This figure also shows that the classical domain of the  steepest descents wave function is bounded below by a critical value of $\p0 \sim {\cal O}(1)$, because there are no regular saddle points associated with classical histories for smaller values of $\p0$. 

%%%%%%%%%%%%%%%%%%%%%%%%%%
%%%%%%%%%%%%%%%%%%%%%%%%%

Our results for the transition probabilities $p_{\rm trans}(\p0'',\p0')$ between two inflationary histories are summarized in Figures
\ref{fig:actionschi1} and \ref{fig:transitionifochi2}. Applied to the classical NBWF ensemble this yields a quantum connection between both ensembles and in particular a small probability to bounce even in the low $\p0$ regime of phase space where the classical extrapolation of the individual asymptotic histories backwards in time produces a singularity. We illustrate this in Fig. \ref{fig:noboundaryWeight}(b) where we compare the $(\p0'',\p0')$ combinations for the most probable quantum transitions with the combinations resulting from a classical extrapolation through the bounce or into a singularity. For large scalar field values on the boundaries the classical and quantum evolution essentially agree. Both predict a symmetric, regular, approximately classical bounce. This is expected because the classicality conditions only fail in a very narrow regime around the bounce in this regime. By contrast, for low values of the initial scalar field the quantum evolution yields a small probability to (quantum) bounce whereas the classical evolution produces a singularity.

\begin{figure}[ht!]
	\centering
	\includegraphics[width=0.75\textwidth]{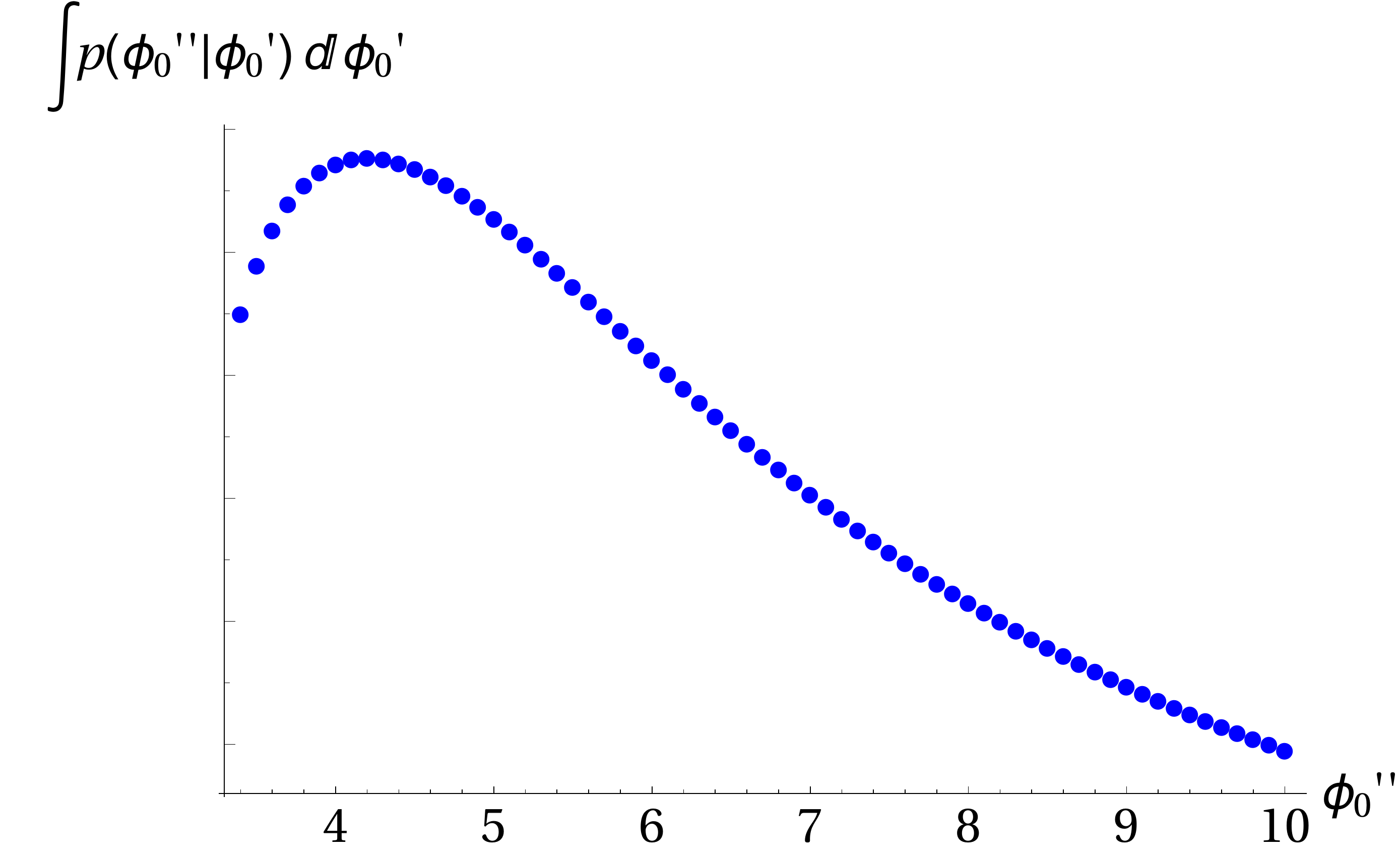}\\
	\caption{The probability distribution over asymptotically classical no-boundary histories labeled by $\p0''$ that exhibit a quantum bounce, marginalized over the histories on the other side of the bounce labeled by $\p0'$.}
	\label{fig:weightingsA}
\end{figure}

From the two dimensional probability distribution \eqref{probhist2} of bouncing histories with no-boundary conditions on both ends one can construct a marginal distribution by integrating over $\p0'$. This yields a distribution over $\p0''$ that can be viewed as a particular branch of the past evolution of the usual NBWF histories in which one conditions on there being a bounce in the past. This distribution is shown in Fig.\ref{fig:weightingsA}. The competition between the no-boundary weighting favoring low $\p0''$ and the transition probabilities favoring high $\p0''$ yields a most probable history at an intermediate value of $\p0''$. 

%%%%%%%%%%%%%%%%%%%%
%%%%%%%%%%%%%%%%%%%%

\begin{figure}[ht!]
	\centering
	\includegraphics[width=0.45\textwidth]{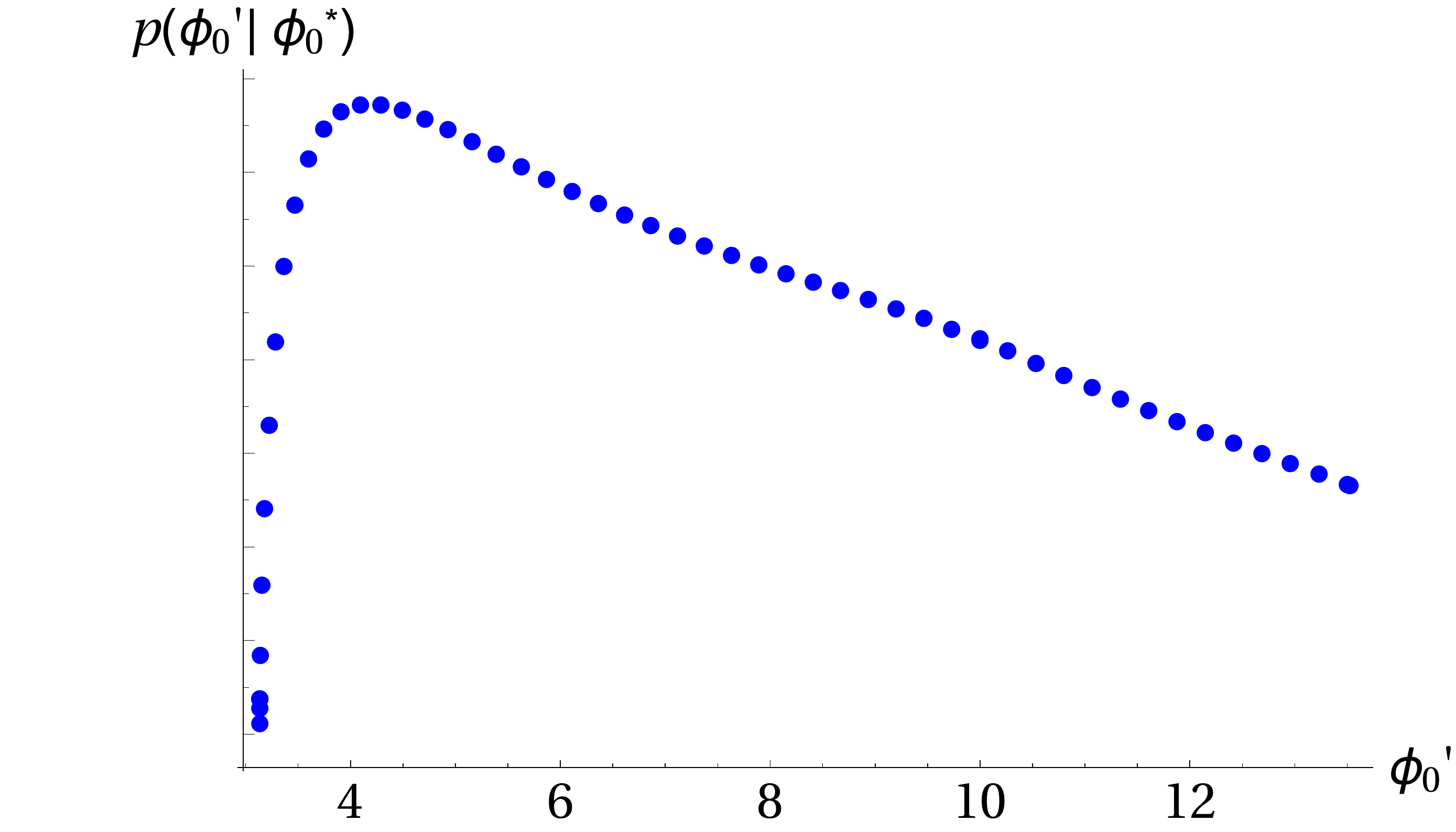}
	\includegraphics[width=0.45\textwidth]{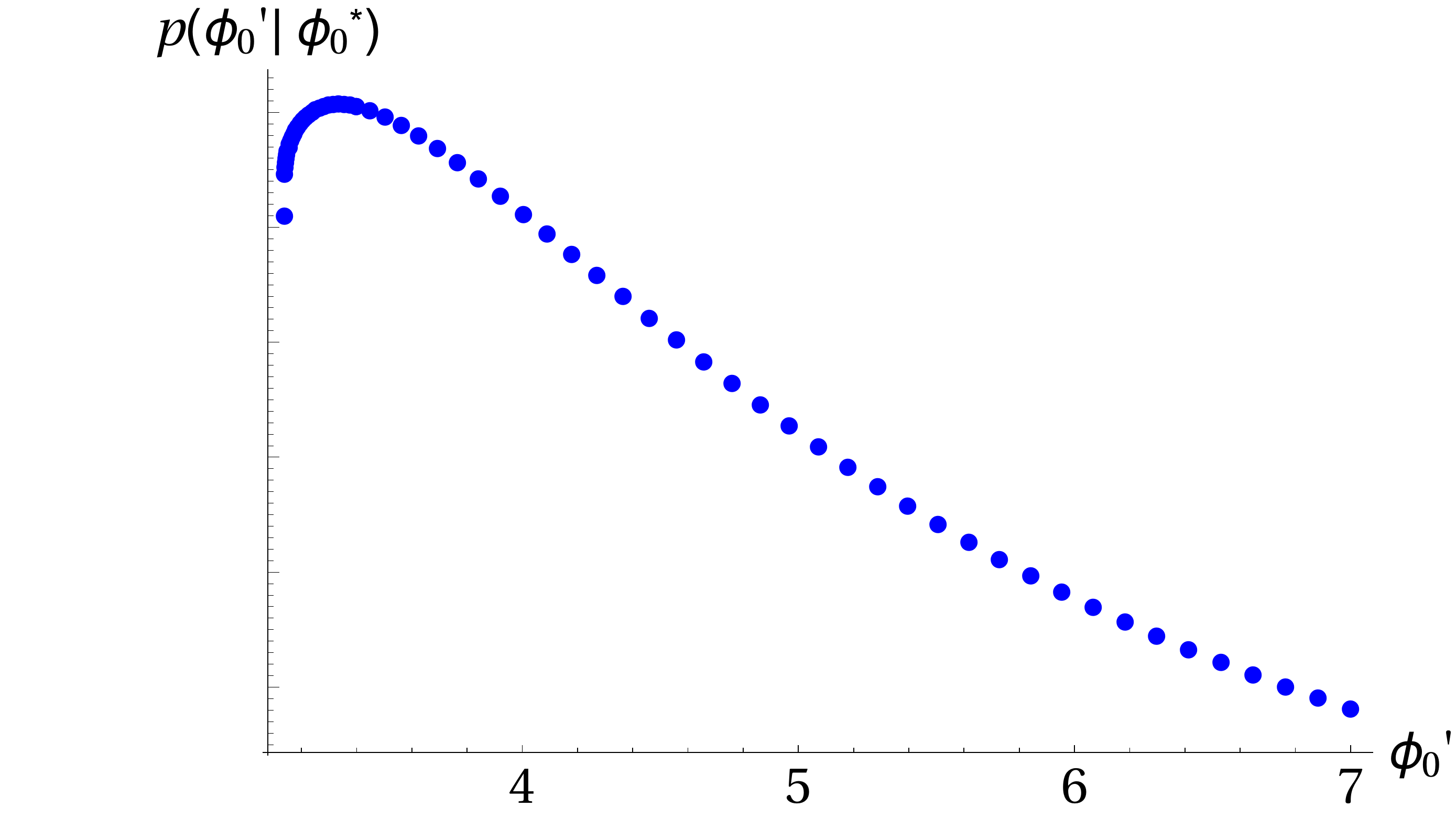}\\
	\caption{Logplots for the conditional probability \eqref{bncpast} that our present history labeled by $\p0^*$ arose from a quantum transition from an earlier classical history $\p0'$, for two different values of $\phi_0''$. In the left panel $\p0^*=10.0$ and in the right panel $\p0^*=3.4$.}
	\label{fig:weightings}
\end{figure}

From the probabilities for entire histories \eqref{probhist2} one can also construct various conditional probabilities. For instance, suppose by measurements of the expansion we determine that we find ourselves in a late time classical history labeled by $\p0''=\p0^*$. What is the probability, with initial and final no-boundary conditions, that this present history arose from a quantum transition from an earlier classical history $\p0'$? This is given by the conditional probability
\be
\label{bncpast}
p(\p0'|\p0^*) \equiv p(\p0',\p0^*)/p(\p0^*) = \ptr(\p0^*,\p0')\pnb(\p0') \ . 
\ee
We show this in Fig. \ref{fig:weightings} for two values of $\p0^*$.

%%%%%%%%%%%%%%%%%%%%
%%%%%%%%%%%%%%%%%%%%

However, the past evolution on the opposite side of a bounce is unlikely to lead to testable predictions on our side of the bounce. This is because the physical arrows of time point away from the bounce on both sides, consistent with overall time symmetry \cite{Hartle:2011rb}. To have an effect on our observations, signals from events on the far side would have to propagate backwards in the time direction defined by the thermodynamic arrow there and then through a quantum epoch. But the situation is radically different in ekpyrotic cosmology to which we turn next.

%%%%%%%%%%%%%%%%%%%
\section{Quantum Transitions: from Ekpyrosis to Inflation}\label{section:ekinf}
%%%%%%%%%%%%%%%%%%%

The methods we have developed are not confined to the case of inflationary dynamics. In fact, the inflationary case is slightly special in that the transitions relate classical histories with opposite arrows of time. As a separate road to pursue, one may also consider big crunch singularities, and ask whether it is possible to tunnel out of them into an expanding universe, thereby avoiding the big crunch. The best understood example of a big crunch is that of an ekpyrotic phase, which is a phase of high-pressure contraction during which anisotropies are suppressed \cite{Khoury:2001wf,Lehners:2008vx}. Thus, during such a phase the universe is driven towards a spatially flat crunch, and this justifies our minisuperspace approach. 

We should note that various models for transitions from the contracting into an expanding phase have been investigated to date: in the original ekpyrotic model, the big crunch was modelled as the collision of higher-dimensional branes \cite{Khoury:2001wf,Khoury:2001bz}. At the classical level, the crunch was still singular (even though the singularity is much milder from a higher-dimensional point of view \cite{Lehners:2006pu}), and thus the precise evolution across such a transition rests on assumptions of how to match a contracting with an expanding universe across a singular surface, see e.g. \cite{Cartier:2003jz,Tolley:2003nx}. To improve the calculational reliability, non-singular bouncing models were also constructed \cite{Buchbinder:2007ad,Creminelli:2007aq,Lehners:2011kr,Ijjas:2016vtq} (for an implementation within the NBWF see \cite{Lehners:2015efa}). Such models have the great advantage that one can calculate explicitly and unambiguously what happens to the background evolution and to cosmological perturbations (and it was found, for instance, that long-wavelength cosmological perturbations evolve across the bounce without being altered \cite{Koehn:2013upa,Battarra:2014tga}). However, all of the currently known models include hypothetical forms of matter, such as ghost condensates \cite{Creminelli:2006xe} or Galileons \cite{Qiu:2011cy,Easson:2011zy}, with unusual properties and no clear origin in fundamental physics. 

Here we will be concerned with a more direct, and in fact more conservative approach: namely we want to see if one can transition out of an ekpyrotic contraction phase via a quantum transition\footnote{See e.g. \cite{Craps:2007ch,Hertog:2005hu,Barbon:2011ta,Awad:2009bh,Battarra:2010id} for some earlier work on the quantum resolution of cosmological singularities mostly in the context of the holographic approach to quantum gravity.}. As we will demonstrate, this is indeed possible, and in the particular example that we have studied, the ekpyrotic universe performs a quantum transition into an expanding inflationary phase. The reason for transitioning to inflation rather than, say, a kinetic dominated phase, is that the inflationary attractor guarantees a transition to another phase of classical evolution.

\begin{figure}[h] 
		\includegraphics[width=0.5\textwidth]{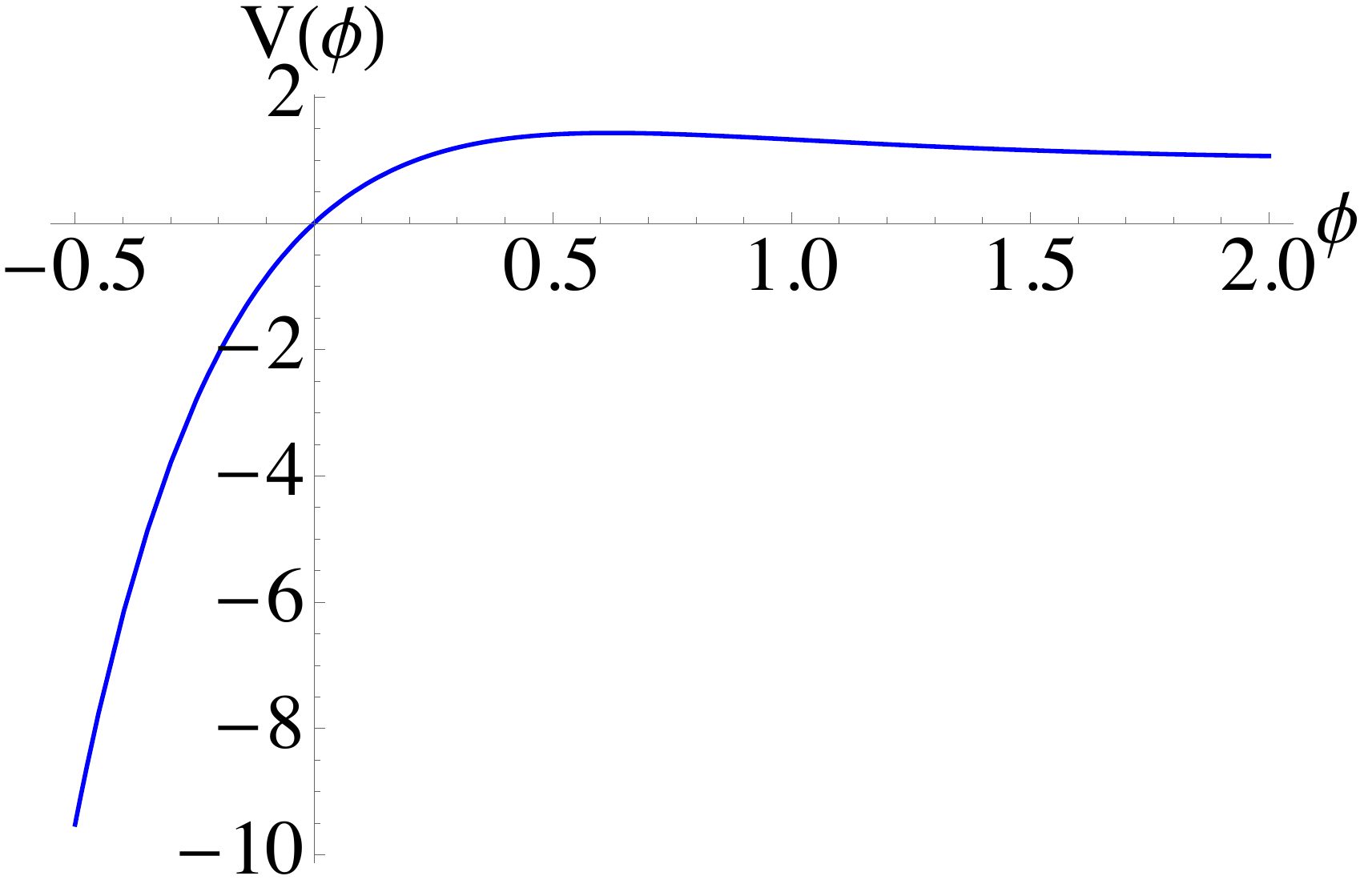}
	\caption{The potential $V(\phi) = 3 \left( 1- e^{-3\phi}\right) + 2 \tanh(-\phi)$ that we are considering in this section. It allows for ekpyrotic contracting solutions on the left, and for inflationary expanding solutions to the right of the maximum. We wish to show that quantum transitions between these two types of solutions are possible.}
	\label{fig:ek_inf_potential}
\end{figure} 

The model that we study again contains gravity coupled to a scalar field with a potential. We take the potential to be of the form
\begin{equation}
V(\phi) = V_0 \left( 1- e^{-c\phi}\right) + 2 \tanh(-\phi) \ ,
\end{equation}
with the constants chosen to be $V_0=c=3.$ The potential is shown in Fig. \ref{fig:ek_inf_potential}. It contains a steep negative region for negative values of $\phi,$ separated by a barrier from a region where the potential is positive and flat for positive $\phi.$ This potential allows for two types of attractor solutions: inflationary slow-roll solutions at positive $\phi$, and ekpyrotic contracting solutions at negative $\phi$. Let us be slightly more specific about the ekpyrotic solutions: for $\phi \lesssim -1,$ we can approximate $V(\phi) \approx -3e^{-3\phi}.$ Assuming a standard flat Robertson-Walker metric $\mathrm{d}s^2 = - \mathrm{d}t^2 + a^2(t) \mathrm{d}\bf{x}^2,$ this model allows for the scaling solutions \cite{Battarra:2014kga}
\begin{eqnarray}
\label{eq:ekpyroticattractor1}
a(t) & = & a_0 (-t) ^{1/ \epsilon} \left( 1 + \frac{ \sqrt{2 \epsilon}}{3} \,  \alpha \, (-t) ^{1- 3/ \epsilon}  + \ldots - \frac{1- 3 \epsilon}{3(1- \epsilon)} \frac{1}{ \sqrt{2 \epsilon}} \beta \,(-t) ^{2-2/ \epsilon} + \ldots \right) \;, \\ \label{eq:ekpyroticattractor2}
\phi(t) & = & \sqrt{ \frac{2}{ \epsilon}} \ln \left( - \sqrt{ \frac{ \epsilon^2  V_0}{ \epsilon- 3 }} t \right)  + \alpha \, (- t) ^{1 - 3/ \epsilon}  + \ldots + \beta \, (-t) ^{2 - 2/ \epsilon} + \ldots \;,
\end{eqnarray} 
where $a_0$ is a constant and where we have included the leading correction terms. The parameters $\alpha, \beta$ are fixed by initial conditions. Here $\epsilon = c^2/2 = 9/2$ is the fast-roll parameter, which by definition is always larger than $3$ during an ekpyrotic phase. This expression clearly shows that the scaling solution is an attractor, as all correction terms die off in the approach to an eventual big crunch at $t=0.$ Note that the approximation of a spatially flat metric is justified since both the energy of expansion $H^2 \propto t^{-2} \propto a^{-2\epsilon}=a^{-9}$ and the energy density of the scalar field $\dot\phi^2 \propto a^{-9}$ grow much faster than the energy density in anisotropic fluctuations (which scales as $a^{-6}$) as the universe contracts. 

\begin{figure}[h] 
		\includegraphics[width=\textwidth]{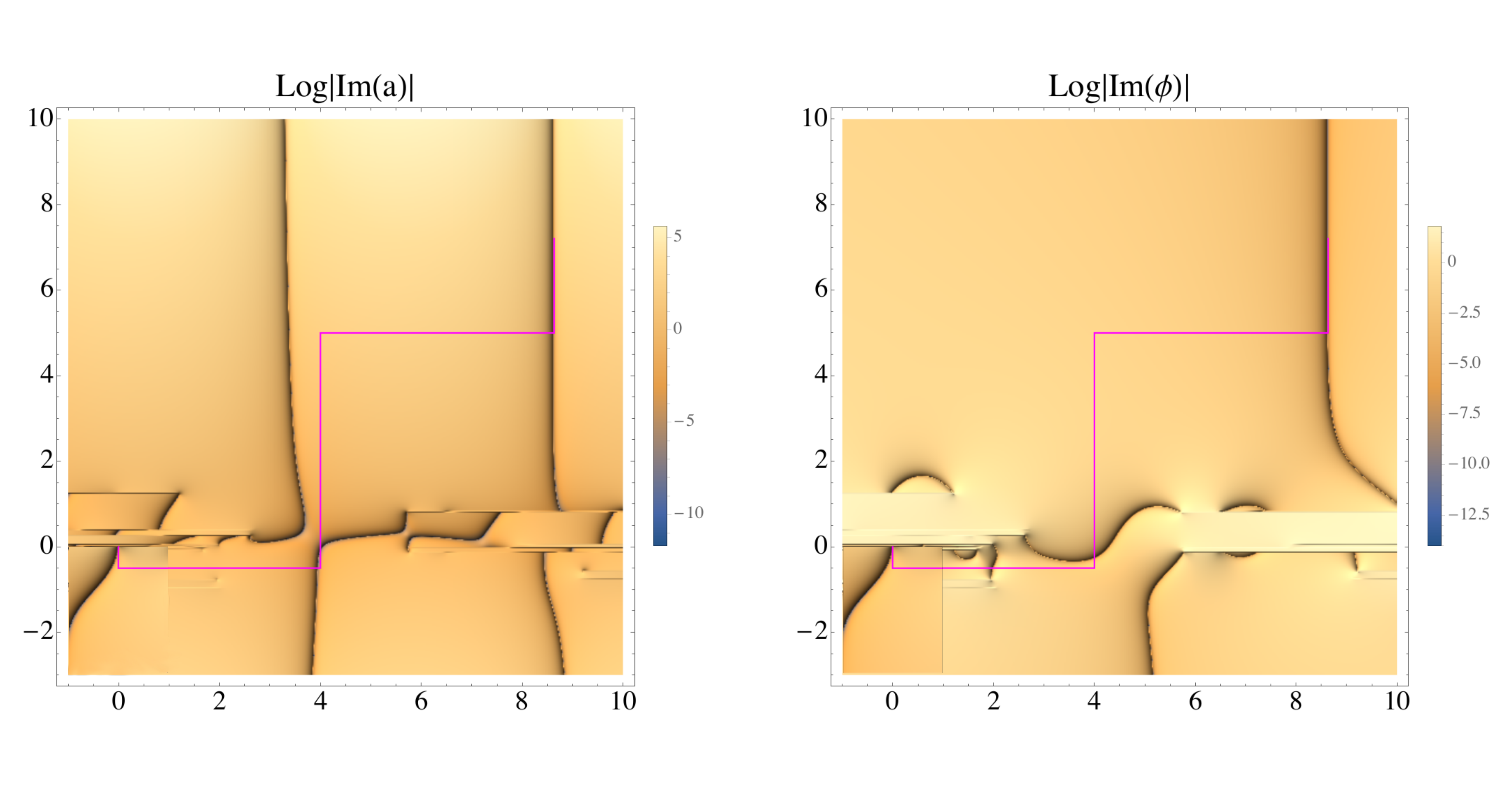}
	\caption{These relief plots show our solution interpolating between a contracting ekpyrotic phase and an expanding inflationary phase over a region of the complex time plane. More specifically, the plots show (the logarithm of the absolute value of) the imaginary part of the scale factor (left panel) and scalar field (right panel), with darker colours corresponding to smaller imaginary parts. Thus the dark lines show the locus where the fields take real values. The bottom left part of the figures show an ekpyrotic history headed for a big crunch at $t=0,$ while the upper right part shows the final inflationary history with coincident lines of real scale factor and scalar field. On the ekpyrotic side, the lines of real $a$ and $\phi$ also become coincident in the approach to the crunch, but this occurs over a very small time interval just before the crunch -- this is as expected from studies of ekpyrotic instantons \cite{Battarra:2014xoa}. The graph necessarily only shows one sheet of the full solution function, while one can clearly distinguish several singular points and the associated branch cuts.}
	\label{fig:ek_inf_relief}
\end{figure} 

\begin{figure}[h] 
		\includegraphics[width=0.75\textwidth]{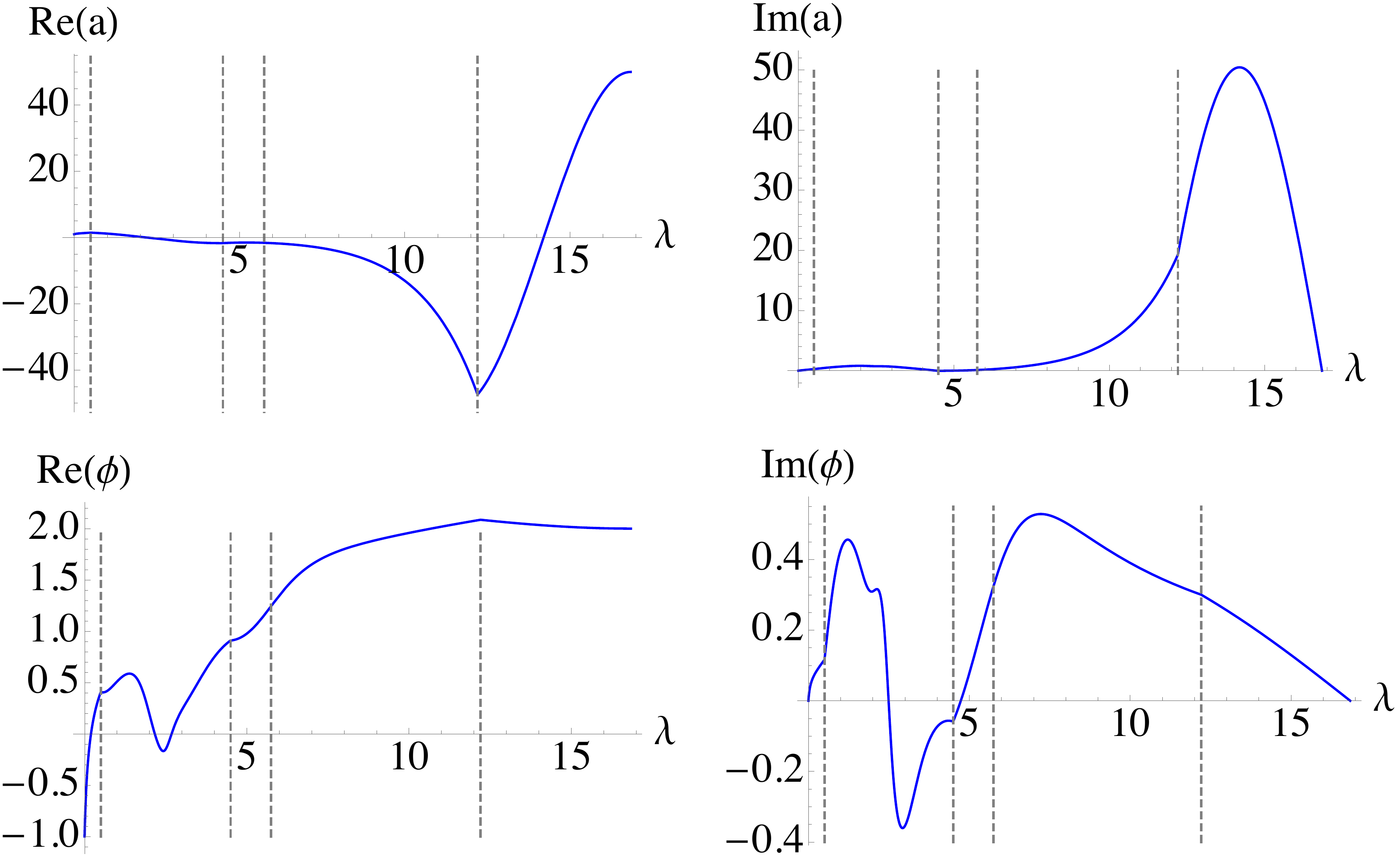}
	\caption{These graphs show the evolution of the scale factor $a$and the scalar field $\phi$ along the contour (parameterised by $\lambda$) shown by a pink line in Fig. \ref{fig:ek_inf_relief}. Note that the fields are real at the end points, as required. In fact, for this particular solutions, we have $b_1=1,\, \chi_1=-1$ and $b_2=50,\, \chi_2=2.$ The solution corresponds to imposing $\chi_{1,\tau} = -1.40549 +  14.1652 i$ and $b_{1,\tau} = -0.582476 + 
 5.69670 i ,$ with $b_{1,\tau}$ being determined by the Hamiltonian constraint. At small $\lambda,$ the contour first runs down, and accordingly along this first segment the fields undergo a reverse ekpyrotic contraction. Along the middle segments, the evolution is fully complex, and along the final vertical segment a real inflationary expanding history has been reached. At large $\lambda,$ it is also obvious from the plots of the action that the imaginary part varies fast compared to the variation in the real part, which shows that the WKB classicality conditions will be satisfied there. More prosaically, this can already be guessed from the fact that the imaginary parts of $a$ and $\phi$ are tiny there, compared to the real parts. At small $\lambda,$ it is less obvious that the ekpyrotic starting history is indeed classical in a WKB sense, which is why we have performed a more detailed WKB analysis as shown in Fig. \ref{fig:ek_inf_WKB}.}
	\label{fig:ek_inf_path}
\end{figure} 

The question now is whether, while classically headed for disaster, the big crunch can be avoided by a quantum transition to the other attractor solution, namely the inflationary one at positive scalar field values. Semi-classically, such a quantum transition can be described by a complex saddle point of the path integral, i.e. we will once again look for a complex solution of the equations of motion, this time interpolating between an ekpyrotic starting point and an inflationary final point. Finding such a solution is complicated in this case by the presence of numerous singularities, which arise because along the ekpyrotic part of the potential a singularity can be reached within a finite time. Because of these singularities, it is not obvious what the appropriate contour in the complex time plane ought to be, and some trial and error is inevitable. An example of an interpolating solution is shown in Fig. \ref{fig:ek_inf_relief}, and the evolution of the fields along the contour drawn as a pink line in Fig. \ref{fig:ek_inf_relief} is shown in Fig. \ref{fig:ek_inf_path}. The many singularities mentioned above are immediately apparent in Fig. \ref{fig:ek_inf_relief}, which shows only the relevant sheet of the solution function. Encircling the singularities in different ways typically leads to an entirely different solution, usually containing no region of classicality. This solution is a showcase example of the use of complex time paths in describing quantum tunneling, as described in more detail in \cite{Turok:2013dfa,Bramberger:2016yog}.

\begin{figure}[h] 
\begin{minipage}{0.5\textwidth}
		\includegraphics[width=0.9\textwidth]{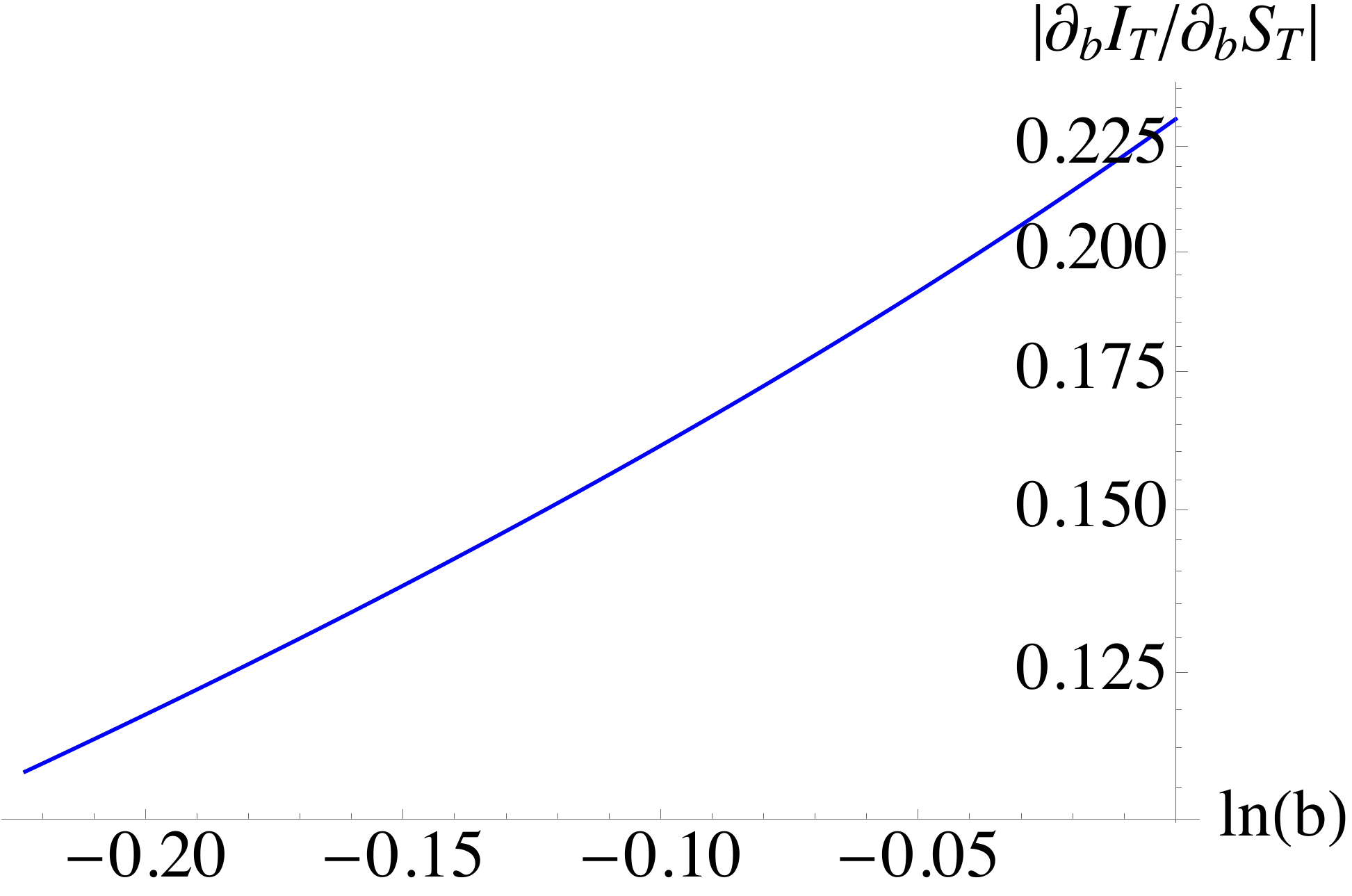}
	\end{minipage}%
	\begin{minipage}{0.5\textwidth}
		\includegraphics[width=0.9\textwidth]{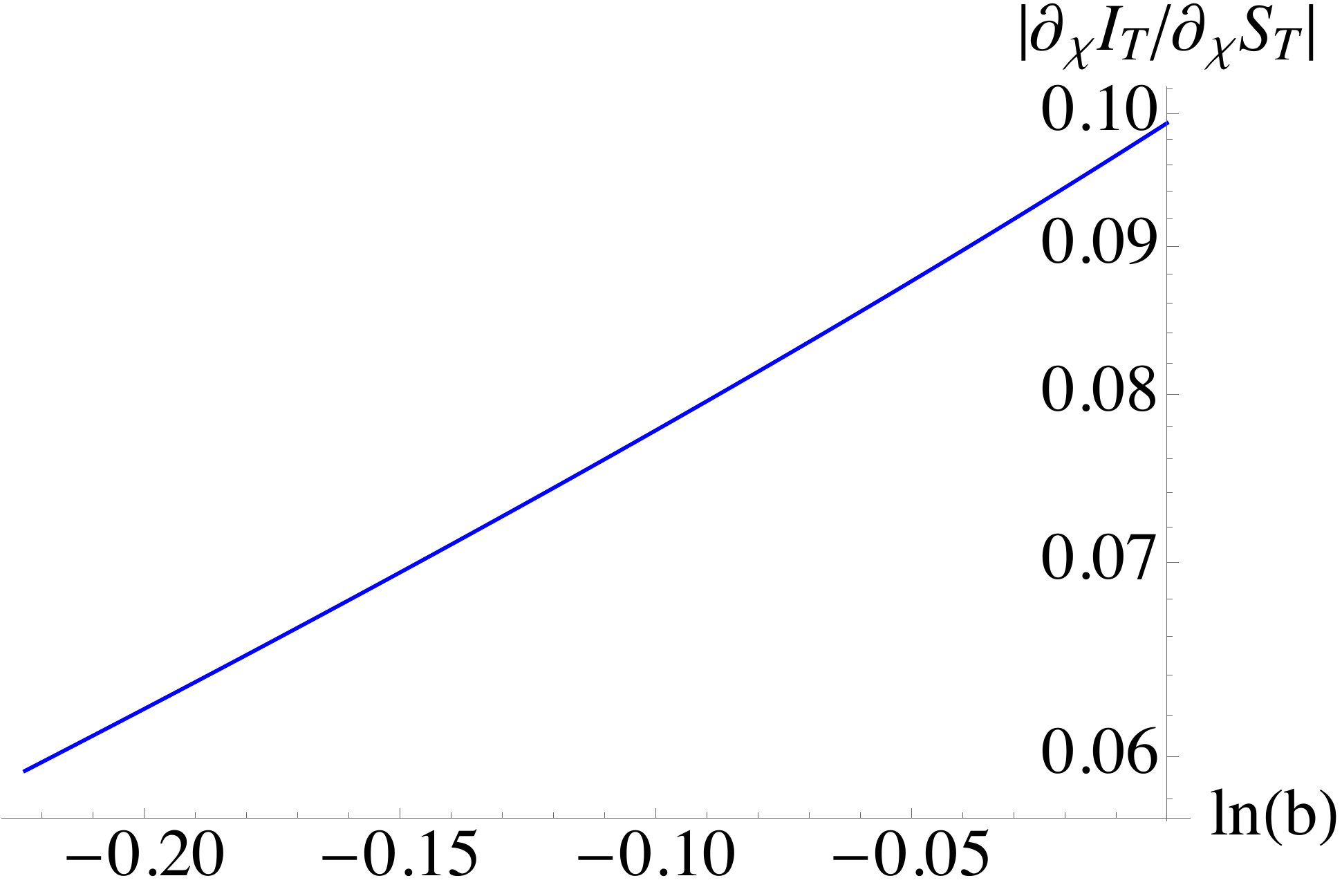}
	\end{minipage}%
	\caption{A graph of the WKB classicality conditions \eqref{classcond}. These become small as the universe contracts, i.e. as $b$ decreases, thus demonstrating that the approach to a big crunch is accurately described by a classical ekpyrotic phase.}
	\label{fig:ek_inf_WKB}
\end{figure}

As one can see from the figures, our solution indeed interpolates between an ekpyrotic and an inflationary history, with a fully complex evolution in between. It is obvious that a classical inflationary solution is reached near the final boundary, since the scale factor and scalar field remain real in the Lorentzian direction for an extended period of time. On the ekpyrotic side, although at the starting point the fields are real, they rather quickly develop imaginary parts too. This is because the ekpyrotic contraction occurs over a very short time period, and thus appears compressed in the figure. In order to show unambiguously that we are indeed starting from a classical ekpyrotic contraction history, we have evaluated the WKB classicality conditions \eqref{classcond} on the ekpyrotic side, keeping the final field values on the inflationary end fixed (while allowing the field derivatives to vary).  Corresponding plots are shown in Fig. \ref{fig:ek_inf_WKB}. As the figures show, the WKB conditions are better and better satisfied as the universe contracts towards a big crunch, which is just as expected for the ekpyrotic attractor \cite{Battarra:2014xoa,Lehners:2015sia}. This result also implies that the probability for tunnelling out of an ekpyrotic phase is constant along a classical ekpyrotic contracting solution, analogously to the inflationary case treated in section \ref{inftoinf}. 

The solution that we have just presented may be considered as a proof of principle that quantum transitions out of an ekpyrotic contracting phase and into an expanding inflationary phase are possible.

%%%%%%%%%%%%%%%%%%%
\section{Discussion} \label{discussion}
%%%%%%%%%%%%%%%%%%%

We have shown that classical cosmological singularities can be resolved in minisuperspace semi-classical quantum gravity and replaced by quantum bounces interpolating between contracting and expanding branches of cosmological histories. 

We have focussed on two cases of special interest where classical cosmological evolution often involves a singularity: transitions from inflation to inflation, and transitions from ekpyrosis to inflation. Let us summarise our findings. 

The quantum bounces that we have found are mediated by complex saddle points of the action of gravity coupled to a scalar field, interpolating between specified real initial and final classical configurations. For the case of inflationary-to-inflationary transitions, the symmetry of the problem selects the appropriate contour of integration in the complex plane. This provides a clean starting point to identify more general instantons describing asymmetric transitions obtained by smoothly deforming away from the symmetric case. Interestingly, at large values of the inflaton potential, the most likely transition turns out to be the symmetric one whereas at low values of the potential -- the regime where the classical extrapolation produces a singularity -- tunnelling to a slightly larger value of the potential is preferred. 

Combined with the no-boundary wave function, which provides a measure on inflationary cosmology, our results for the quantum bounces yield probabilities for an ensemble of complete inflationary histories exhibiting a quantum transition that connects two classical inflationary patches on either side. The quantum transitions identified here allow one to refine and differentiate between different possible pasts of the inflationary histories in the NBWF, which are coarse grained over in the usual treatment. 

We have also analysed a potential that contains both ekpyrotic contracting and inflationary expanding solutions. In this case, we have demonstrated the existence of similar quantum transitions from the contracting phase into the expanding one, avoiding the big crunch singularity that in a purely classical context would follow the ekpyrotic contraction. For the ekpyrotic-to-inflationary transitions we do not have any guidelines as to what the appropriate contour of integration should be, and thus we do not know yet whether the solutions that we have found are the dominant ones. An important question for future work is to clarify this question, perhaps by generalising the treatment of quantum mechanical tunnelling described in \cite{Bramberger:2016yog} to include gravity. Another question is whether one can transition to other phases of classical evolution, such as a radiation or matter dominated universe.

Our treatment of quantum transitions has close connections with two papers that recently appeared: in \cite{Gielen:2015uaa} the authors considered a radiation-dominated universe in a theory with a conformal coupling to gravity, which allows for technically very simple complex instantons characterised by $a \propto \eta,$ where $\eta$ denotes the conformal time. Such instantons avoid the big crunch at $\eta=0$ by choosing a contour that passes around the singularity. Note that these solutions necessarily interpolate between negative and positive values of the scale factor, which appears to be a doubling of phase space in the models we considered in this paper. And in \cite{Chen:2016ask} the authors found approximate complex instantons interpolating between two phases of de Sitter expansion in the more restrictive context of gravity coupled to a cosmological constant. 

Beyond the questions alluded to above, our work opens up the possibility to place ekpyrotic cosmology on firmer footing. Specifically it provides a starting point to address the central open question of the evolution of perturbations through a classically singular bounce, and possible observational signatures of the pre-bounce era. This will require a generalisation of our treatment that includes cosmological perturbations. We leave this question for upcoming work.

\acknowledgments
We thank James Hartle, George Lavrelashvili and Neil Turok for stimulating discussions over many years. The work of SFB is supported in part by a grant from the Studienstiftung des Deutschen Volkes. The work of TH and YV is supported in part by the National Science Foundation of Belgium (FWO) grant G092617N, by the C16/16/005 grant of the KULeuven and by the European Research Council grant no. ERC-2013-CoG 616732 HoloQosmos.

%%%%%%%%%%%%%%%%%%%%%%%%%%%%%%%%%%%%%
\appendix

\section{Contours of Integration} \label{contour}

\begin{figure}[ht!]
	\centering
	\begin{minipage}{0.9\textwidth}
	\includegraphics[width=0.9\textwidth]{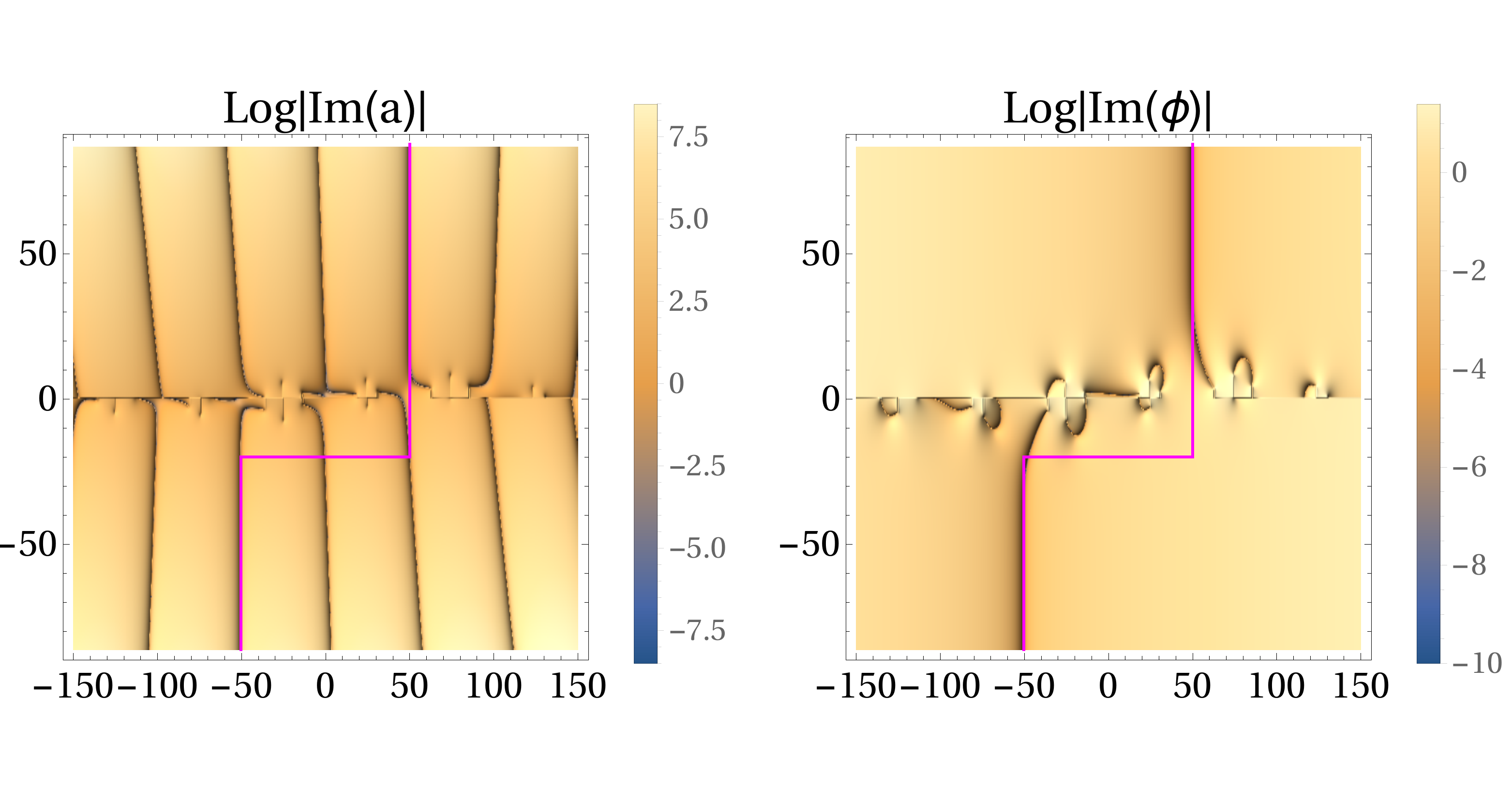}
	\end{minipage}	
	\caption{An instanton with an L-shaped path connecting $\chi_1=10$ to $\chi_2=10.$}
	\label{fig:imvaluesLBounce}
\end{figure}

\begin{figure}[ht!]
	\centering
	\includegraphics[width=0.75\textwidth]{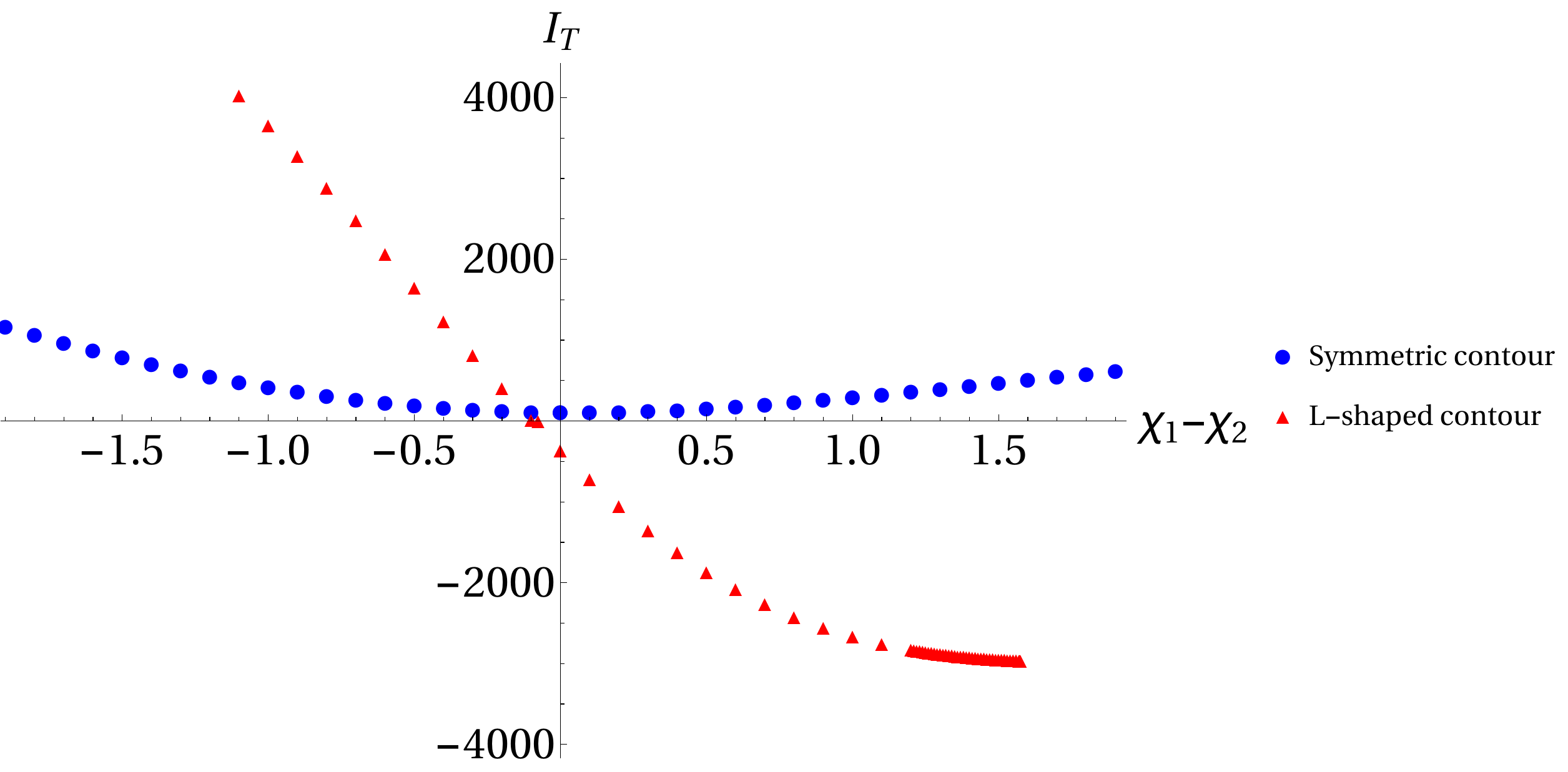}
	\caption{The real part of the action for transition from $\chi_1=10$ to various values of $\chi_2,$ plotted as a function of $\chi_1-\chi_2$ and for two types of integration contour: the L-shaped one (in blue dots) and the symmetric one (in red squares). }
	\label{fig:actionsBandL}
\end{figure}

For inflationary-to-inflationary transitions, apart from the specific symmetric contour used in the main part of the text, other paths are possible which also lead to potentially valid interpolating solutions. One example is an L-shaped path, as shown in Fig. \ref{fig:imvaluesLBounce}. From the final classical history, this path runs straight down through the point where the big bang singularity would have been, had the solution been real. The fact that the solution is complex now allows one to continue through this point, and connect with an incoming classical history in the bottom left quadrant. However, as shown in Fig. \ref{fig:actionsBandL}, if we look at transitions to different scalar field values, then for the L-shaped path the implied probability distribution would be non-normalisable, indicating that this class of solutions may not be physical. Similar results are obtained for other L-shaped paths that connect with further-removed loci of real $a$ values, and with upside-down L-shaped paths that run through the would-be singularity from the bottom up. For this reason we focus on the symmetric contour. 

\begin{figure}[ht!]
	\centering
	\includegraphics[width=0.75\textwidth]{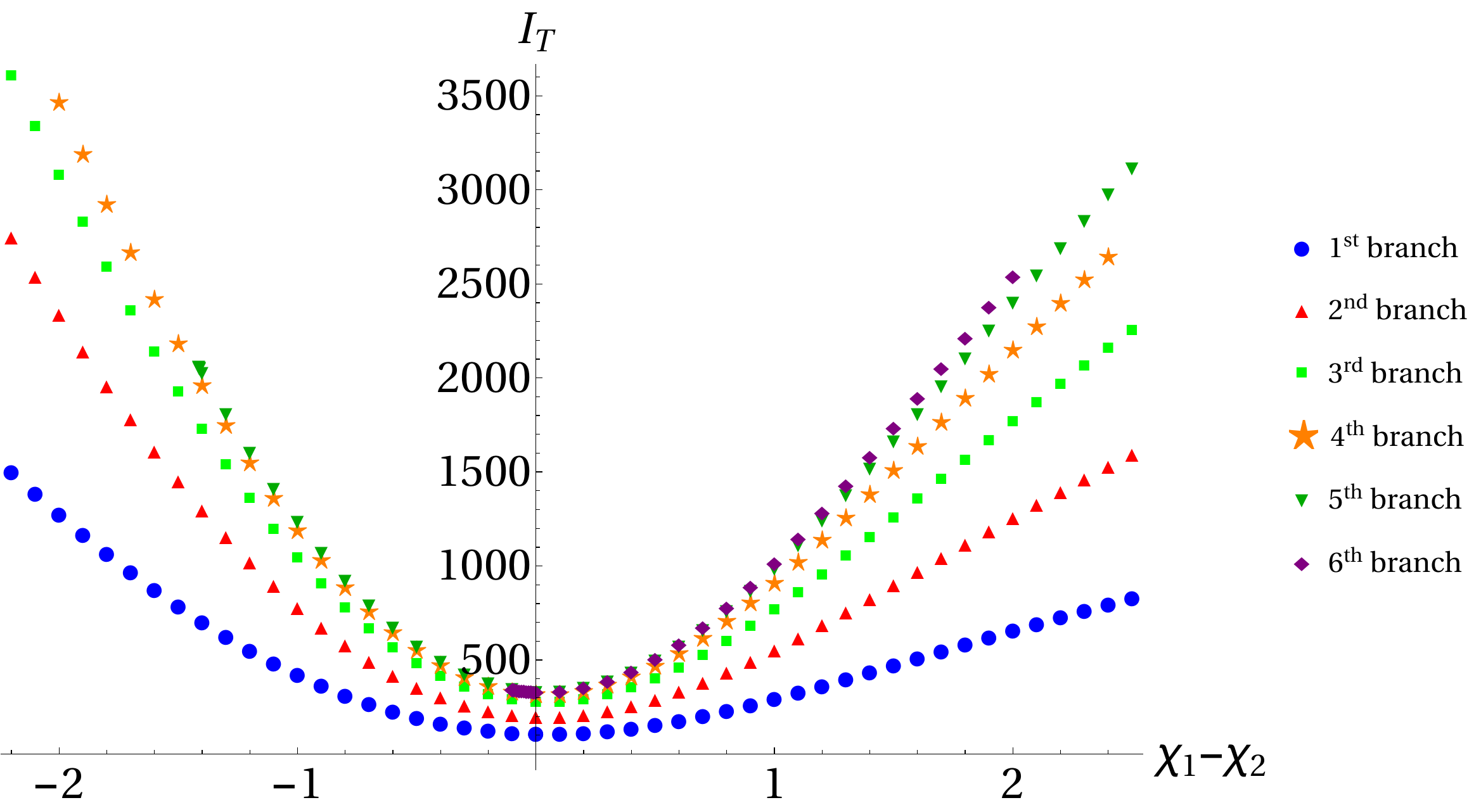}
	\caption{The real part of the action for symmetric contours connecting classical histories that are ever further separated in Euclidean time, and which we refer to as different branches. These solutions have higher actions and are thus further suppressed.}
	\label{fig:actionsSymBs}
\end{figure}

But even for the symmetric contour, we have further possibilities, as we can use it to connect classical histories that are further separated in Euclidean time. Fig. \ref{fig:actionsSymBs} shows the real part of the action for such transitions between increasingly separated ``branches'' where the scale factor is real, each time with field derivatives optimised such that the locus of real scalar field asymptotically overlaps with the line of real scale factor. For these higher branches, the action increases monotonically, indicating that these transitions, though also normalisable, are further suppressed. We may thus safely ignore these.

\section{Perturbative Results} \label{analytic}

For inflationary-to-inflationary transitions, the equations of motion can be solved analytically in various approximate regimes, allowing us to provide approximate analytic descriptions of the quantum transitions. 

\subsection{Large scalar field}
At large field values we will be in the slow-roll regime, ${\phi'}^2 \ll V(\phi)$ and $\phi'' \ll \frac{\partial V(\phi)}{\partial \phi}$. If the scale factor is also large, then the spatial curvature can be ignored and the equations of motion simplify to
\begin{subequations}
\begin{align}
3\frac{a'}{a} &= m^2 \frac{\phi}{\phi'} \ , \label{eqn:modeomratio} \\
a''& +\frac{a \kappa^2 m^2 \phi^2}{6}=0 \ , \\
\left(\frac{a'}{a}\right)^2 &= -\frac{\kappa^2 m^2 \phi^2}{6} \ .
\end{align}
These can be solved explicitly, for example by substituting the first equation into the last,
\begin{align}
{\phi'}^2= -\frac23 \frac{m^2}{\kappa^2} \ ,
\end{align}
\end{subequations}
which can be easily solved to give
\begin{align}
\phi(\tau) = \phi(0) \pm i \frac{m}{\kappa} \sqrt{\frac23}\tau \ .
\end{align}
Notice that again we took the point of symmetry $\tau_s=0$.
This result for $\phi$ can be plugged into \eqref{eqn:modeomratio}, from which we then obtain the sale factor
\begin{align}
\frac{a'}{a}=\mp i\frac{\kappa m}{\sqrt{6}} \left( \phi(0) \pm i \sqrt{\frac23}\frac{m \tau}{\kappa}\right) \Rightarrow a(\tau)= C e^{\mp i \frac{\kappa m \phi(0) \tau}{\sqrt{6}} + \frac{m^2 \tau^2}{6}} \ .
\end{align}
To find the value of $C$, we will follow Lyons \cite{Lyons:1992ua}, who states that when Re$(\phi(0))>0$ in the upper half $\tau-$plane the solutions with the upper sign are valid while in the lower half $\tau$-plane the ones with the lower sign are valid. This means that around $\tau = 0$ both solutions should be matched to one solution, 
\begin{align}
\phi(\tau)& \approx \phi(0) \ ,\\
a(\tau) & \approx C \cos\left(\frac{\kappa m \phi(0)}{\sqrt{6}} \tau\right) \ .\label{eqn:soltausmall}
\end{align}
Notice that it is clear that these solutions obey the bounce boundary conditions that we impose. To find $C$ we plug $a$ into the Hamiltonian constraint, giving $C= \sqrt{\frac32 }\frac{1}{\kappa m \phi(0)}$. Collecting our results, the solution is approximately given by
\begin{align}
\phi(\tau)&= \phi(0) + i \sqrt{\frac23}\frac{m \tau}{\kappa} \ , \\
a(\tau) & = \sqrt{\frac32} \frac{1}{\kappa m \phi(0)} \exp\left(-i\frac{\kappa m \phi(0)}{\sqrt{6}} \tau + \frac{m^2 \tau^2}{6}\right) \ .
\end{align}

From this we can understand the behaviour in our numerics. It is clear that $\phi$ is real along $\tau= X_{TP} + i t$ with $X_{TP}= -\phi_{sI} \frac{\kappa}{m} \sqrt{\frac23}$. Here we have split  $\phi(0)$ in its real and imaginary components  $\phi(0) = \phi_{sR} + i \phi_{sI}$. It is also interesting to write $\phi(0)=\phi_s e^{i \theta_s}$, from which we can see that a change in phase, keeping $\phi_s$ fixed, shifts the line of real $\phi(\tau)$:
\begin{align}
X_{TP}= - \phi_s \sin(\theta_s) \sqrt{\frac32} \frac{\kappa}{m} \ .
\end{align}
Let us now take a closer look at the scale factor. To see where it becomes real, we will try to write it as an amplitude times a phase,
\begin{align}
a(\tau)= & \sqrt{\frac32}\frac{1}{\kappa m \phi_s} \exp\left[\frac{m^2}{6}(x^2-t^2) +\frac{\kappa m \phi_s}{\sqrt{6}}( \cos (\theta_s) t + \sin(\theta_s) x) \right] \nonumber \\
& \times \exp\left[i \left(\frac{m^2 x t}{3} -\theta_s- \frac{\kappa m \phi_s}{\sqrt{6}}(\cos(\theta_s) x - \sin(\theta_s) t) \right) \right] \ . \label{eqn:solaabsphase}
\end{align}
This becomes real when the phase is a multiple of $\pi$. If $t$ is not too big, the only relevant term is the one without $t$, therefore we expect that $a$ is real along
\begin{align}
X= \frac{n \pi \sqrt{6}}{\kappa m \phi_s \cos(\theta_s)} \ , \label{eqn:xtps}
\end{align}
with $n \in \mathbb{Z} $. This explains why we see different lines in the complex $\tau-$plane along which $a$ is real. In the numerical results there are singularities around $\tau=0$, though these do not appear in our analytic results. The lines of real $a$ and $\phi$ will coincide if
\begin{align}
-\phi_{sI} \frac{\kappa}{m} \sqrt{\frac32}= \frac{n \pi \sqrt{6}}{\kappa m \phi_{sR}} \Rightarrow \phi_{sI}= -\frac{2 n \pi}{\kappa^2 \phi_{sR}} \ ,
\end{align}
or written in terms of the absolute value and the phase of $\phi(0)$
\begin{align}
\theta_s=\frac12 \sin^{-1}\left(\frac{4 n \pi}{\kappa^2 \phi_s^2}\right) \ .
\end{align}
Thus we expect no obstruction to finding such interpolating solutions numerically.

\subsection{Small scalar field}
Another region where analytic results are possible is the region where the scalar field is a small perturbation, without backreaction on the metric. Starting with the case for which there is no scalar field at all, the equation of motion for $a$ is solved by
\begin{align}
a''(\tau)=0 \Rightarrow a(\tau)= A + B \tau \ .
\end{align}
Because we demand $a'(0)=0$ we can conclude that without a scalar field the scale factor is constant, $a(\tau) = A$. If we now add a small scalar field to this background, we get an equation of motion for $\phi$:
\begin{align}
\phi'' -m^2 \phi=0 \ . 
\end{align}
The solution for this, obeying the appropriate boundary condition $\phi'(0)=0$, is
\begin{align}
\phi(\tau)= \phi_s e^{i \theta_s} \cosh(m \tau) \ .
\end{align}
Plugging this together with $a$ into the Hamiltonian constraint gives  $a(\tau)=\pm \frac{\sqrt{6}}{ \kappa m \phi_s e^{i \theta_s}}$.

Now we can look for regions where both $a$ and $\phi$ are real. The only values of $\theta_s$ for which $a$ is real are $\theta=n \pi$. $\phi$ can then only be real along the imaginary axis, i.e. $X_{TP}=0$. This explains why we can not find complex bounce solutions for very small $\phi_s$, the only possible solutions are those that are real.

\bibliography{Literature}

\end{document}